\documentclass[12pt]{article}
\usepackage[a4paper, margin=1.2in]{geometry}

\usepackage{amssymb}
\usepackage{latexsym}
\usepackage{amsmath}
\usepackage{amsthm}
\usepackage{appendix}
\usepackage{graphicx}
\usepackage{wrapfig}
\usepackage{authblk}
\usepackage{epstopdf}

\usepackage{natbib}
\usepackage{algpseudocode}
\setcitestyle{authoryear,open={(},close={)}}

\def\mrm{\ensuremath\mathrm}
\def\mbf{\ensuremath\boldsymbol}

\newcommand{\cov}[1]{\mrm{cov}(#1)}

\usepackage{url}
\usepackage[svgnames]{xcolor}

\usepackage[colorlinks]{hyperref}
\hypersetup{citecolor=DarkBlue, linkcolor=Blue, urlcolor=Blue}

\def\anonymized{0}

\newcommand{\anon}[2]{
\if\anonymized1
#1
\else
#2
\fi
}

\begin{document}

\title{An organ deformation model using Bayesian inference to combine population and patient-specific data}

\anon{}{
\author[1,2]{Øyvind Lunde Rørtveit\thanks{Corresponding author. oyvind.rortveit@uib.no}}
\author[1,2]{Liv Bolstad Hysing}
\author[3,2]{Andreas Størksen Stordal}
\author[1,2]{Sara Pilskog}
\affil[1]{Haukeland University Hospital, Bergen, Norway}
\affil[2]{University of Bergen, Norway}
\affil[3]{NORCE Norwegian Research Centre, Bergen, Norway}
}
\renewcommand\Affilfont{\itshape\small}
\maketitle

\begin{abstract}
\noindent\textbf{Objective:}\indent 
Organ deformation models have the potential to improve delivery and reduce toxicity of radiotherapy, but existing data-driven motion models are based on either patient-specific or population data. We propose to combine population and patient-specific data using a Bayesian framework. Our goal is to accurately predict individual motion patterns while using fewer scans than previous models.\\
\noindent{\textbf{Approach:}}\indent
We have derived and evaluated two Bayesian deformation models. The models were applied retrospectively to the rectal wall from a cohort of prostate cancer patients. These patients had repeat CT scans evenly acquired throughout radiotherapy.  Each model was used to create coverage probability matrices (CPMs). The spatial correlations between these CPMs and ``true'' CPMs, derived from independent scans of the same patient, were calculated.\\
\noindent{\textbf{Main results:}} \indent
Spatial correlation with ground truth were significantly higher for the Bayesian deformation models than both patient-specific and population-derived models with 1, 2 or 3 patient-specific scans as input. Statistical motion simulations indicate that this result will also hold for more than 3 scans. \\
\noindent{\textbf{Significance:}}\indent
The improvement over known models means that fewer scans per patient are needed to achieve accurate deformation predictions. The models have applications in robust radiotherapy planning and evaluation, among others.

\end{abstract}

%\maketitle

\section{Introduction}
In radiotherapy (RT), the dose is carefully shaped to the patient anatomy as seen in the CT acquired before start of treatment (plan CT),  to achieve a good compromise 
between disease control and risk of inducing complications.  Since the variability of the organ positions and deformations is unknown before start of treatment, different measures have been adopted to safeguard against motion uncertainties through planning margins \citep{stroom_inclusion_1999,van_herk_probability_2000}, robust optimization \citep{unkelbach_robust_2018} and/or treatment plan adaptation \citep{yan_adaptive_1997}. 

A statistical model for the deformation of organs of individual patients using principal component analysis (PCA) of the organ's surface shape vectors was first proposed by \cite{sohn_modelling_2005}. The main drawback of the patient-specific model is that the number of data samples (in the form of organ contours derived from 3D images) per patient is often low, which limits the robustness of the motion estimates \citep{thornqvist_adaptive_2013}. 

\citet{budiarto_population-based_2011} proposed a population based statistical model, under the assumption that, although the size, shape and position of organs differ greatly between patients, the patterns of deformation are generally the same. The advantage is that an estimate of a patient's deformation patterns exists even when only a single observation is available. When applied to prostate target deformation, they showed that about 50\% of the variation could be explained by 15 population deformation modes (i.e. principal components). Subsequent uses of the population model include \citet{bondar_statistical_2014}, who used it to create margins for rectal cancer patients, \citet{rios_population_2017}, who modeled bladder deformation for prostate cancer RT, \citet{szeto_population_2017} who modeled daily variations in the thorax, and \citet{magallon-baro_modeling_2019}, who modeled deformation in the stomach, duodenum and bowel for pancreatic cancer RT. A weakness of the population model is its inability to model patient-specific deformation patterns, even when multiple scans are available for the patient in question. The aim of the current work is to combine the strengths of the population and patient-specific models by  introducing Bayesian models that take in to account both the population deformation patterns (in terms of a \emph{prior} distribution) and patient-specific measurements, forming an individualized \emph{posterior} distribution. Bayesian models have previously been applied to the problem rigid shifts of the patient, termed setup errors \citep{lam_application_2005,herschtal_finding_2012}.

In this paper, we introduce two Bayesian models, which differ in their choice of priors. The choice of model to use will be a trade-off between accuracy and simplicity. We derive necessary algorithms to efficiently calculate the approximate posterior distributions in high dimensions. We apply the introduced models to a realistic example with complex motion, in terms of the rectal wall of prostate cancer patients. We use the models to estimate coverage probability matrices (CPMs), i.e. 3D-arrays of voxels where the value in each voxel is the probability that the voxel will be covered by the rectal wall at any given time. We compare the accuracy of CPMs estimated using the two Bayesian methods, the patient-specific model by \citet{sohn_modelling_2005} and the population model by \citet{budiarto_population-based_2011}. 

In addition to the presentation of new models, this is to our knowledge the first comparison between these two previous models, as well as the first time such an organ deformation model has been applied to the rectum.

\section{Methods}
\begin{figure}[h]
\centering
\includegraphics[width=0.25\textwidth]{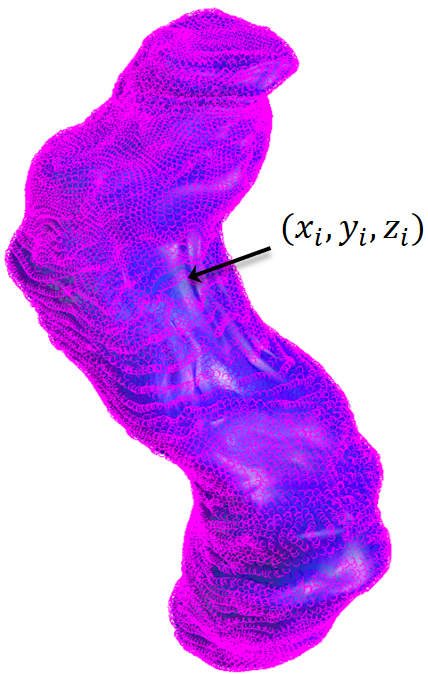}
\caption{A rectum shape represented by a set of organ surface points.}
\label{xyzfigure}
\end{figure}

In the class of deformation models that we study, an organ shape is represented by a set of points on the organ surface, as illustrated in figure \ref{xyzfigure}. These representations are derived from organ contours segmented from 3D images; for simplicity we refer to one organ shape of this kind as a ``scan''. The $x$, $y$ and $z$ coordinates of all $P$ points are gathered into a \emph{shape vector} $s$:
\begin{equation}
s =[x_1, y_1, z_1, x_2, y_2,z_2,\ldots,x_P, y_P, z_P]^T.
\end{equation}
With this representation, we can use standard multivariate statistical distributions. 

To compare organs across scans, we need corresponding points between all shapes in the data set. This correspondence is found using deformable and rigid contour registration both within and between patients.  Details are beyond the scope of the current work, but can be found in\anon{(citation anonymized)}{\citet{rortveit_reducing_2021}}.

An assumption for all the following methods is that, for a specific patient, the shape coordinates follow a multivariate Gaussian distribution:
\begin{equation}\label{sconditional}
s\sim\mathcal{N}(\mu, R).
\end{equation}
The mean shape vector $\mu$ represents the patient's mean organ shape, and the covariance matrix $R$ describes the variance of the coordinates as well as the covariance between each pair of coordinates. When $\mu$ and $R$ are given, we can use the distribution to draw new random organ shapes for the patient. 

The difference between the methods is how $\mu$ and $R$ are estimated. In the Bayesian methods introduced in section \ref{Bayesianmodels} , $\mu$ and $R$ are considered random samples from specific prior distributions, whose parameters are calculated from the training data. \emph{Point estimates} of $\mu$ and $R$ are derived from the posterior distributions.

Due to the high dimensions of the shape vectors, all covariance matrices are parametrized using principal component analysis (PCA), see e.g. \citet[chapter~10]{fujikoshi_multivariate_2010}. Under PCA, a covariance matrix is represented by a few eigenvectors and corresponding eigenvalues. These are usually found through singular value decomposition (SVD) of a \emph{data matrix} $D$, whose columns are normalized mean-subtracted samples, such that $R=DD^T$. 

\subsection{Patient-specific model}
In the patient-specific model introduced by \cite{sohn_modelling_2005}, only data from the patient under consideration is used. The mean shape is thus the average of the available shapes $s_1, s_2, \ldots, s_J$ for that patient;
\begin{equation}\label{indivmeancalc}
\hat{\mu} = \bar{s}=\frac{1}{J}\sum_{j=1}^J s_j,
\end{equation}
while $R$ is the sample covariance matrix:
\begin{equation}
\hat{R}_{\mrm{ps}} = \frac{1}{J-1}\sum_{j=1}^J(s_j-\hat{\mu})(s_j-\hat{\mu})^T.
\end{equation}
\subsection{Population model}
The population model introduced by \cite{budiarto_population-based_2011}  rests on the assumption that the covariance matrix is the same for all patients, and only the mean differs. The mean is calculated as the mean shape vector for the individual patient as in \eqref{indivmeancalc}, while the covariance matrix is the average of the sample covariance matrices for each patient in the training set:
\begin{equation}
\hat{R}_\mathrm{pop}=\frac{1}{M}\sum_{i=1}^M \hat{R}_i=\frac{1}{M}\sum_{i=1}^M\frac{1}{J_i-1}\sum_{j = 1}^{J_i}(s_{i,j}-\bar{s}_i)(s_{i,j}-\bar{s}_i)^T.\label{Rpop}
\end{equation}
\subsection{Bayesian models}
\label{Bayesianmodels}
In Bayesian inference, new data is combined with prior knowledge (such as population statistics) to update probability distributions. In the following, the mean and covariance matrix for a given patient are considered random parameters that vary across the population according to a prior distribution, $f(\mu, R)$. When data for a new patient is available, we can compute the posterior distribution $f(\mu, R|\mbf{s})$, where $\mbf{s}=\{s_1, s_2,\ldots,s_J\}$, through Bayes theorem:
\begin{equation}
f(\mu, R|\mbf{s})= \frac{f(\mbf{s}|\mu, R)f(\mu, R)}{f(\mbf{s})}\label{generalposterior}
\end{equation}
Bayes theorem gives us a distribution of the possible values of $\mu$ and R, as opposed to single values. Nevertheless, due to the complexity of the posterior distributions in our subject matter, we shall resort to looking at point estimates of $\mu$ and $R$, such as the expected value or mode of the posterior. 

The Bayesian models we present differ in the selection of the prior density. We resort to priors that result in computationally feasible posterior distributions, since Markov Chain-Monte Carlo methods are computationally expensive in high dimensions. In the following sections, we present two priors which each represent a Bayesian model.

\subsubsection{Normal-Inverse-Wishart prior}
A natural way of combining the estimates of the covariance matrix from the patient-specific model and population model is to use a simple weighted average:
\begin{equation}
\hat{R}=\lambda \hat{R}_\mrm{pop}+(1-\lambda)\hat{R}_\mrm{ps}
\end{equation}
The weighting should be proportional to the number of samples used to compute the estimates. By setting $\lambda = \frac{\nu}{\nu + J}$ for some $\nu$, we obtain
\begin{equation}
\hat{R}=\frac{1}{\nu+J} (\nu\hat{R}_\mrm{pop}+J\hat{R}_\mrm{ps}),
\end{equation}
which is the same result as obtained from assuming an inverse Wishart (IW) prior for $R$ and using a specific point estimate for $\hat{R}$ for the posterior, as shown below. 

IW is a matrix distribution, and a \emph{conjugate prior} to the multivariate Gaussian likelihood with known mean and unknown covariance matrix. This means that the posterior distribution for $R$ is also IW, and the parameters are obtained from equations involving the prior parameters and the data. The parameters of the IW are the \emph{scale matrix} $\Psi$ and the \emph{degrees of freedom} $\nu$. Formally, if $\mu$ is given, and the prior for $R$ is IW,
\begin{equation}
R \sim \mathcal{IW}(\Psi, \nu),
\end{equation}and the likelihood is Gaussian,
\begin{equation}
s|R\sim \mathcal{N}(\mu, R),
\end{equation}
then the posterior $R|\mbf{s}$, where $\mbf{s}=\{s_1,s_2,\ldots,s_J\}$ is also IW,
\begin{equation}
R|\mbf{s}\sim \mathcal{IW}(\Psi^\prime, \nu^\prime),
\end{equation}
with posterior parameters
\begin{align}
    \Psi^\prime& = \Psi + \sum_{j=1}^J (s_j-\mu)(s_j-\mu)^T\label{PsiprimeIW}\\
    \nu^\prime &= \nu + J.\label{nuprimeIW}
\end{align}
We define $\Psi=\nu\hat{R}_\mrm{pop}$ and use the point estimate $\hat{R} =\frac{1}{\nu^\prime}\Psi^\prime$ for the posterior. Inserting both these expressions into \eqref{PsiprimeIW}, we get
\begin{equation}
\hat{R} = \frac{1}{\nu+J}\left(\nu\hat{R}_\mrm{pop}+\sum_{j=1}^J (s_j-\mu)(s_j-\mu)^T\right)\label{RupdateIW}
\end{equation}
The parameter $\nu$ determines the weight between the population covariance matrix and the sample covariance matrix of the new patient, and can be selected either by tuning or by optimization. One can think of $\nu$ as encoding the strength of our belief that $\hat{R}_\mrm{pop}$ can represent our new patient's covariance matrix.

In reality, $\mu$ is not given. One could replace $\mu$ by $\hat{\mu}$ from \eqref{indivmeancalc}, but this will lead to bias in the covariance matrix estimate when $J$ is small (to see this, consider equation \eqref{RupdateIW} when $J=1$ and therefore $\hat{\mu}=s_1$). Instead, we consider both $\mu$ and $R$ random, and look for a joint prior distribution.

The conjugate prior for the multivariate Gaussian likelihood with both unknown mean and covariance is the Normal-Inverse-Wishart (NIW) distribution. In the NIW, $R$ is IW-distributed, i. e. 
\begin{equation}
R \sim \mathcal{IW}(\Psi, \nu),\label{RIW}
\end{equation}
but $\mu$ and $R$ are not independent. The conditional distribution of $\mu$ given $R$ is Gaussian:
\begin{equation}\label{mugivenR}
\mu|R\sim \mathcal{N}(\mu_0, \frac{1}{\kappa} R).
\end{equation}
Thus, the NIW has the parameters $\mu_0$, $\kappa$, $\Psi$ and $\nu$, and we write
\begin{equation}
\mu, R\sim\mathcal{NIW}(\mu_0, \kappa, \Psi, \nu).
\end{equation}Here, $\mu_0$ is the population mean, and the scalar $\kappa$ represents the ratio of the variance between scans of the same patient (intra-patient) to the variance between patients (inter-patient). 

Since this is a conjugate prior, the posterior is also NIW, and we can write
\begin{equation}
\mu, R|\mbf{s}\sim\mathcal{NIW}(\mu_0^\prime, \kappa^\prime, \Psi^\prime, \nu^\prime)
\end{equation}
with \citep{bishop_pattern_2006}
\begin{align}
\mu_0^\prime &= \frac{1}{\kappa + J}(\kappa\mu_0 + J\bar{s})\label{muUpdateNIW}\\
\kappa^\prime& = \kappa + J\\
\nu^\prime &= \nu + J\\
\Psi^\prime& = \Psi + \sum_{j=1}^J (s_j-\bar{s})(s_j-\bar{s})^T+\frac{\kappa J }{\kappa + J}(\bar{s}-\mu_0)(\bar{s}-\mu_0)^T.\label{PsiprimeNIW}
\end{align}
Note the similarity between \eqref{muUpdateNIW} and \eqref{RupdateIW}: Both are weighted averages between population and patient-specific estimates, with the weight of the patient-specific estimate proportional to the number of patient-specific samples $J$. Hence, both $\nu$ and $\kappa$ are parameters which determine the weight between the population and patient-specific estimates. 

The final term of \eqref{PsiprimeNIW} can be considered a correction for the uncertainty of the sample mean, which makes the equation different from \eqref{PsiprimeIW}, where the mean was assumed to be known. 

The maximum a-posteriori (MAP) estimate of $\mu$ is the expected value of the posterior, $\mu_0^\prime$, so we let
\begin{equation}
\hat{\mu}=\frac{1}{\kappa + J}(\kappa\mu_0 + J\bar{s}).\label{muHatNIW}
\end{equation}When only a single observation for the new patient is available, i. e. $J=1$, \eqref{muHatNIW} becomes identical to the shrinkage estimation from \cite{rortveit_reducing_2021}.

As for the IW-case, we let $\Psi=\nu \hat{R}_\mrm{pop}$ and $\hat{R}=\frac{1}{\nu^\prime}\Psi^\prime.$  Inserting this into \eqref{PsiprimeNIW} yields
\begin{align}
\hat{R} = &\frac{1}{\nu+J}\Bigg(\nu \hat{R}_\mrm{pop}+\sum_{j=1}^J (s_j-\bar{s})(s_j-\bar{s})^T \nonumber\\
&  +\frac{\kappa J}{\kappa + J}(\bar{s}-\mu_0)(\bar{s}-\mu_0)^T\Bigg).
\end{align}
In practice, we never construct the full covariance matrix $\hat{R}$. Instead, it is represented by a data matrix which is augmented with extra columns, such that $D^\prime D^{\prime T}=\hat{R}$. Given the population data matrix $D$, where $DD^T=\hat{R}_\mrm{pop}$, and the patient-specific data matrix $S$ whose columns are $s_j-\bar{s}$ for $j=1\ldots J$, the augmented data matrix is
\begin{equation}
    D^\prime =\frac{1}{\sqrt{\nu+J}}\left[\sqrt{\nu}D\quad\sqrt{\frac{kJ}{k+J}}(\bar{s}-\mu_0)\quad S\right].
\end{equation}

\subsubsection{Variational Bayes model}
In the NIW-model, the covariance matrix of the patient mean $\mu$ is proportional to $R$, as seen in \eqref{mugivenR}. But the assumption that the intra-patient covariance is proportional to the inter-patient covariance may in practice not be fulfilled. A more flexible approach is to separate the two, which motivates the following model.

Assume that the mean $\mu$ is Gaussian distributed according to 
\begin{equation}
\mu \sim \mathcal{N}(\mu_0, \Lambda).\label{muNormal}
\end{equation}
Here, $\mu_0$ is the population mean, and $\Lambda$ is the \emph{inter-patient covariance matrix} that describes how the individual mean varies from patient to patient. Assume further that $R$ is IW distributed according to \eqref{RIW}, and $\mu$ and $R$ are independent; i.e.
\begin{equation}
f(\mu, R) = \mathcal{N}(\mu;\mu_0, \Lambda)\cdot \mathcal{IW}(R;\Psi, \nu).
\end{equation}
Unfortunately, this prior is not conjugate to the Gaussian likelihood \eqref{sconditional}, and there is no simple expression for the posterior. However, both $\mu$ and $R$ follow tractable posterior distributions \emph{when conditioned on the other}, namely
\begin{equation}\label{muigivenRi}
\mu|R, \mbf{s}= \mathcal{N}(\mu_0^\prime, \Lambda^\prime)
\end{equation}
and 
\begin{equation}\label{Rigivenmui}
R|\mu, \mbf{s}= \mathcal{IW}(\Psi^\prime, \nu^\prime).
\end{equation}
Prior distributions with this property are said to be \emph{conditionally conjugate} to the likelihood. The conditional posterior parameters $\mu^\prime$, $\Lambda^\prime$, $\Psi^\prime$ and $\nu^\prime$ are
\begin{align}
    \mu_0^\prime&=(\Lambda^{-1}+JR^{-1})^{-1}(\Lambda^{-1}\mu_0+JR^{-1}\bar{s})\label{mu0prime}\\
    \Lambda^\prime&=(\Lambda^{-1}+JR^{-1})^{-1}\\
    \Psi^\prime& = \Psi + \sum_{j=1}^J (s_j-\mu)(s_j-\mu)^T\label{Psiprime}\\
    \nu^\prime &= \nu + J.\label{nuprime}
\end{align}
The derivation of \eqref{muigivenRi}-\eqref{nuprime} is given in appendix \ref{app:wishartnormal}. 

Since both $\mu$ and $R$ are unknown, the left hand sides of \eqref{mu0prime}-\eqref{Psiprime} cannot be computed directly from the right hand sides. An alternative is to use an approximative method, known as Mean Field Variational Bayes (MFVB) \citep{gelman_bayesian_1995}. This method is applicable for conditionally conjugate priors, and is a technique used to approximate a complicated posterior distribution by a simpler distribution. The joint posterior distribution of the dependent parameters are approximated by two marginal posterior distributions by assuming independence. In our case, we are looking for densities $q_\mu()$ and $q_R()$ such that
\begin{equation}\label{varbayes}
    q_\mu(\mu)q_R(R)\approx f(\mu, R | S).
\end{equation}
In appendix \ref{app:variational}, we show that $q_\mu$ is a multivariate Gaussian probability density function (pdf), and $q_R$ is an inverse Wishart pdf,
\begin{equation}
f(\mu, R |\mbf{s})\approx N(\mu;\mu_0^*,\Lambda^*)\cdot\mathcal{IW}(R;\Psi^*,\nu^*),
\end{equation}
where the parameters are
\begin{align}
 \Lambda^*&=(\Lambda^{-1}+J\nu^*\Psi^{*-1})^{-1}\label{Lambdaupdate}\\
    \mu_0^*&=\Lambda^*(\Lambda^{-1}\mu_0+J\nu^*\Psi^{*-1}\bar{s})\label{muUpdate}\\
    \Psi^*& = \Psi + \sum_{j=1}^J (s_j-\mu_0^*)(s_j-\mu_0^*)^T+J\Lambda^* \label{PsiUpdate}\\
    \nu^*&=\nu+J.
\end{align}
Equations \eqref{Lambdaupdate}-\eqref{PsiUpdate} must be solved for $\Psi^*$, $\Lambda^*$ and $\mu^*$, but solving them analytically is not possible. We use instead a common iterative technique, where, starting at an initial guess for the parameters, the equations are iterated until convergence. 
If $\Psi^{*(0)}$ is the initial guess for $\Psi^*$, we get the following algorithm:
\begin{algorithmic}
\For{$i=1\ldots$ (until convergence)}
\State $\Lambda^{*(i)}=(\Lambda^{-1}+J\nu^*\Psi^{*(i-1)-1})^{-1}$
\State $\mu_0^{*(i)}=\Lambda^{*(i)}(\Lambda^{-1}\mu_0+J\nu^*\Psi^{*(i-1)-1}\bar{s})$
\State $\Psi^{*(i)} = \Psi + \sum_{j=1}^J(s_j-\mu_0^{*(i)})(s_j-\mu_0^{*(i)})^T+J\Lambda^{*(i)}$
\EndFor
\end{algorithmic}
The iteration is guaranteed to converge to a local optimum, but not necessarily to the global optimum. Whether we find the global optimum or not depends on the starting point. In our case, the prior and the approximate posterior have the same parameters, so the obvious choice of starting point is the corresponding parameter of the prior, i.e $\Psi^{*(0)}=\Psi$. 

Finally, we extract point estimates of $\mu$ and $R$. We let $\hat{\mu}= \mu_0^*$. For the point estimate of $\hat{R}$, see section \ref{sec:deltaR}. Although we are not directly interested in $\Lambda^*$, it is needed in order to calculate the other parameters. $\Lambda^*$ represents the uncertainty about the mean $\mu_0^*$, and as such still contains information that may be valuable depending on application.

Equation \eqref{Lambdaupdate} contains the inversion of 3 matrices, all of which are of dimension $P\times P$. This is not practical; e.g. in our validation data, $P$ is over 50000, so such an inversion would require on the order of $10^{14}$ floating point operations. However, these matrices are highly redundant, as they are estimated from limited data. In practice, we have found that all three update equations \eqref{Lambdaupdate}, \eqref{muUpdate} and \eqref{PsiUpdate} can be computed efficiently without ever constructing any $P \times P$ matrices, and with inversion of much smaller matrices only. The details of the efficient computation are given in appendix \ref{app:computation}.

\subsection{Estimating model parameters from training data}\label{section:training}
Bayesian algorithms require specification of the hyperparameters of the prior. For the present models, these are $\mu_0$, $\kappa$, $\Lambda$, $\Psi$ and $\nu$, with $\kappa$ specific to the NIW-model and $\Lambda$ specific to the variational Bayes model.The vector and matrix valued parameters $\mu_0$, $\Lambda$ and $\Psi$ are estimated from training data. 

Assume that data in the form of shape vectors $s_{i,j}$ from $M$ patients are available, where $i$ is the patient number and $j$ is the scan number, and patient $i$ has $J_i$ scans.

\subsubsection{Population mean}
The prior mean $\mu_0$ is the population mean shape, which is simply calculated as the average of all the individual mean shapes in the training data:
\begin{equation}
    \mu_0=\frac{1}{M}\sum_{i=1}^M\bar{s}_{i}=\frac{1}{M}\sum_{i=1}^M\frac{1}{J_i}\sum_{j=1}^{J_i}s_{i,j}.
\end{equation}

\subsubsection{Population covariance matrix}
\label{psitraining}
The population (or intra-patient) covariance matrix $R_\mrm{pop}$, defined in \eqref{Rpop}, is in practice represented by its principal components and their variances. PCA of such a matrix has been dubbed ``simultaneous component analysis'' (SCA) \citep{timmerman_four_2003}, since all patients are assumed to share the same principal components. The data matrix which is input to SCA contains all the columns from the patient-specific data matrices in the training data:
\begin{equation}
    D_\mrm{pop}=\frac{1}{\sqrt{M}}\left[D_1\quad D_2\quad\ldots\quad D_M\right],
\end{equation}
where $D_i$ is
\begin{equation}
    D_i = \frac{1}{\sqrt{J_i-1}}\left[s_{i,1}-\bar{s}_i\quad s_{i,2}-\bar{s}_i\quad\ldots\quad s_{i,J_i}-\bar{s}\right].
\end{equation}
The covariance matrix $\hat{R}_\mrm{pop} = D_\mrm{pop}D_\mrm{pop}^T$ is used for both the classical population model and the NIW-model. 

In the variational Bayes model, the scale matrix $\Psi$ needs to be invertible. We will use a regularization approach for this model, where we add a constant $\delta_\Psi$ times the identity matrix to the scaled sample covariance matrix:
\begin{equation}
    \Psi=\nu\hat{R}_\mrm{pop}+\delta_\Psi I = \nu D_\mrm{pop}D_\mrm{pop}^T+\delta_\Psi I.
\end{equation}
This structure, together with the similar structure of the inter-patient covariance matrix, makes it possible to compute the update equations\eqref{Lambdaupdate}-\eqref{PsiUpdate} efficiently through the procedure detailed in appendix \ref{app:computation}.

\subsubsection{Inter-patient covariance matrix}
\label{lambdatraining}
In the variational Bayes model, we also need to estimate the covariance matrix $\Lambda$ of $\mu$, which we call the \emph{inter-patient} covariance matrix. This matrix describes how the mean $\mu$ varies between patients.

By definition,
\begin{equation}
    \Lambda = E[(\mu-\mu_0)(\mu-\mu_0)^T].
\end{equation}
We do not have direct observations of $\mu$, but we have estimates, $\bar{s}_i$. A natural extension of the sample covariance matrix suggests an estimator of the form
\begin{equation}
    \hat{\Lambda}_\mathrm{b}=\frac{1}{M-1}\sum_{i=1}^M(\bar{s}_i-\hat{\mu}_0)(\bar{s}_i-\hat{\mu}_0)^T.
\end{equation}
This estimate of $\Lambda$ is biased, since the sample mean $\bar{s}_i$ is not equal to the true mean $\mu$. We show in appendix \ref{app:bias} that the expected value of $\hat{\Lambda}_\mathrm{b}$ is 
\begin{equation}
E[\hat{\Lambda}_\mathrm{b}] = \Lambda +cE[R],
\end{equation}
where $c = \frac{1}{M}\sum_{i=1}^M\frac{1}{J_i}$. The bias is therefore inversely proportional to the number of scans per patient. 

Since $R_\mrm{pop}$ is an unbiased estimate of $E[R]$, we can get an unbiased estimate of $\Lambda$ as
\begin{equation}
    \hat{\Lambda}=\hat{\Lambda}_\mrm{b}-cR_\mrm{pop}.
\end{equation}
However, since both $\hat{\Lambda}$ and $\hat{R}_\mrm{pop}$ are low rank, and they range over different subspaces, the resulting matrix is not positive semidefinite. This makes PCA a bit more complicated, but it is still possible. Details are given in appendix \ref{app:PCAbiascorrected}.

As for the intra-patient covariance matrix, the inter-patient covariance matrix must also be invertible, therefore we add a regularization factor $\delta_\Lambda I$. Additionally, since $\Lambda$ expresses our uncertainly about the mean estimate, we want to have the possibility of increasing its overall size, so we introduce a constant multiplier $\alpha$, which finally leads to
\begin{equation}
    \Lambda = \alpha \hat{\Lambda} + \delta_\Lambda I.
\end{equation}

\subsubsection{Probabilistic PCA}\label{sec:deltaR}
In the NIW-model, we used the point estimate $\frac{1}{\nu^\prime}\Psi^\prime$ for $R$, where $\Psi^\prime$ is the posterior scale matrix, and $\nu^\prime$is the posterior degrees of freedom. In the variational Bayes model, this is less straightforward. The posterior $\Psi^*$ can be expressed as $D^*D^{*T}+\delta^*_\Psi I$ for some $D^*$ and some $\delta^*_\Psi$. The posterior $\delta^*_\Psi$ is approximately proportional to the prior $\delta_\Psi$, and with a large $\delta_\Psi$, the estimate $\frac{1}{\nu^*}\Psi^*=\frac{1}{\nu^*}D^*D^{*T}+\frac{1}{\nu^*}\delta_\Psi^* I$ places an unreasonable amount of variance on the shape coordinates. For this reason, we introduce a new parameter $\delta_R$, and set the point estimate of $R$ to
\begin{equation}
\hat{R}=\frac{1}{\nu^*}D^*D^{*T}+\frac{\nu}{\nu^*}\delta_R I.\label{hatRwithdR}
\end{equation}
For the prior distribution, the point estimate for $R$ is found by replacing the posterior parameters values in \eqref{hatRwithdR} by the equivalent prior parameters. This yields
\begin{equation}
\hat{R}_0 = D_\mrm{pop}D_\mrm{pop}^T+\delta_R I.
\end{equation}
When $D_\mrm{pop}$ is found through PCA, this structure fits the description of \emph{probabilistic PCA} (PPCA) introduced by \cite{tipping_probabilistic_1999}. 
 Their method provides a maximum likelihood estimate for $\delta_R$ given by
 \begin{equation}
 \delta_R = \frac{1}{P-K}\sum_{k = K +1}^P\lambda_k,
 \end{equation}
 where $\lambda_k$ is the $k$th largest eigenvalue of the population covariance matrix in \eqref{Rpop} (i.e. the variance of the $k$th principal component), and $K$ is the number of eigenpairs not discarded in PCA. In other words, $\delta_R$ is the average variance of the discarded dimensions.
 
\section{Evaluation}
\subsection{Material}
For evaluation, we used data from 37 patients with locally advanced prostate cancer. Each patient had 8-10 CT scans taken during treatment (typically 2 per week), in addition to the plan CT used for RT dose planning. No laxatives were administered to the patients before or during treatment. The rectum was defined with content from the recto-sigmoid flexure to the anal verge. One single expert physicist contoured rectum on all CT scans for all patients, and all contours were reviewed and corrected by another expert physicist. This yielded a total of 373 rectum shapes, which were used in leave-one-out cross-evaluation. Details about the patients and treatment can be found in\anon{(citation anonymized)}{\citet{hysing_statistical_2018}}. All shapes from the CT scans were converted to mesh representations with corresponding vertices, using deformable registration. Since toxicity is related to dose to the rectal wall and not its content, we evaluated the methods on the rectal wall. Since the inner wall is not seen on CT scans, we assumed 3 mm wall thickness, as in \citet{sanguineti_refinement_2020}.

\subsection{Parameter values}
The values of the scalar parameters were tuned manually. The values we used are shown in table \ref{table:parameters}. For the parameters $K$-intra and $\nu$, which are applicable to multiple algorithms, we used the same value for all models.

\begin{table}[h]
\centering
\caption{Parameter values for all models. K-intra is the number of principal components used to compute the intra-patient covariance matrix, K-inter is the same for the inter-patient covariance matrix, $\nu$ and $\kappa$ are scalar hyperparameters of the IW/NIW distributions, $\delta_\Lambda$ and $\delta_\Psi$ are regularization parameters for the matrices used in the variational Bayes iteration, and $\alpha$ is the weight of the inter-patient covariance matrix.}
\begin{tabular}{lllllll}\label{table:parameters}
$K$-intra & $K$-inter & $\nu$ & $\kappa$ & $\delta_\Psi$ & $\delta_\Lambda$ & $\alpha$ \\\hline
        12	  &  20      &  6     &    0.25      &     240000        &   80000     &         4
\end{tabular}
\end{table}

\subsection{Coverage probability matrices}
Predicted CPMs were calculated by first estimating $\mu$ and $R$ using the patient-specific, population, and two Bayesian methods, and then generating 500 random rectal wall shapes based on the distribution $\mathcal{N}(\mu, R)$. For each generated shape, we found which voxels (on a $1\times1\times1$ mm grid) were covered by the rectal wall using an in-house developed ray-tracing algorithm. The coverage probability of each voxel was defined as the fraction of generated rectal walls covering that voxel. 

The predicted CPMs and reference CPMs (assumed to be the ``true'' CPMs) were compared in terms of their normalized cross-correlation:
\begin{equation}
c=\frac{\sum_{v \in V}p_\mrm{predict}(v)p_\mrm{true}(v)}{\sqrt{\sum_{v \in V}p_\mrm{predict}^2(v)\sum_{v \in V}p_\mrm{true}^2(v)}},\label{eq:crosscorr}
\end{equation}
where $V$ is the set of all voxels, and $p_\mrm{predict}(v)$ and $p_\mrm{true}(v)$ are the predicted and true coverage probabilities at voxel $v$, respectively. We compared the  results for the different methods using one, two and three input scans. 

We used the remaining independent $J_i-3$ scans for each patient to compute reference CPMs. Since relatively few scans (6-8) were then available, we used the bootstrapping procedure detailed in section \ref{sec:bootstrapped} with this data to generate smooth CPMs. The reference CPM for each patient was computed by drawing 500 bootstrapped rectal wall shapes, and setting the coverage probability of each voxel equal to the proportion of these shapes that covered the voxel.

\subsection{Convergence behaviour}\label{sec:bootstrapped}
 To analyse convergence of the four methods without re-using structures for both training and testing, we created a virtual data set for each patient in the original data set by using a PCA-based bootstrapping procedure: For each patient, we first calculated the principal components using all the patient's available shapes.  We then calculated the PCA-scores for each shape: $c_{i,j,k}$, where $i$ is the patient number, $j$ is the scan number and $k$ is the component number. To generate a new random scan for patient $i$, a new PCA-score $c^*_k$ was drawn for each component number $k$, and a new shape $s_i^*$ was synthesized according to 
\begin{equation}
    s_i^*=\bar{s}_i+\sum_{k=1}^{J_i}c^*_k w_{i,k},
\end{equation}
where  $w_{i,k}$ is the $k$th principal component vector for patient $i$. The $c^*_k$ values were drawn randomly from the existing values $c_{i,j,k}$ for $j=1\ldots J_i$, i.e. by bootstrapping. Since the principal component scores are uncorrelated, such mixing of the scores should create realistic new shapes. The bootstrapping procedure means that no specific distribution has been assumed. 

We used the generated data set to evaluate the model for up to 10 input scans.
\begin{figure*}[h]
    \centering
    \includegraphics[width=1\textwidth]{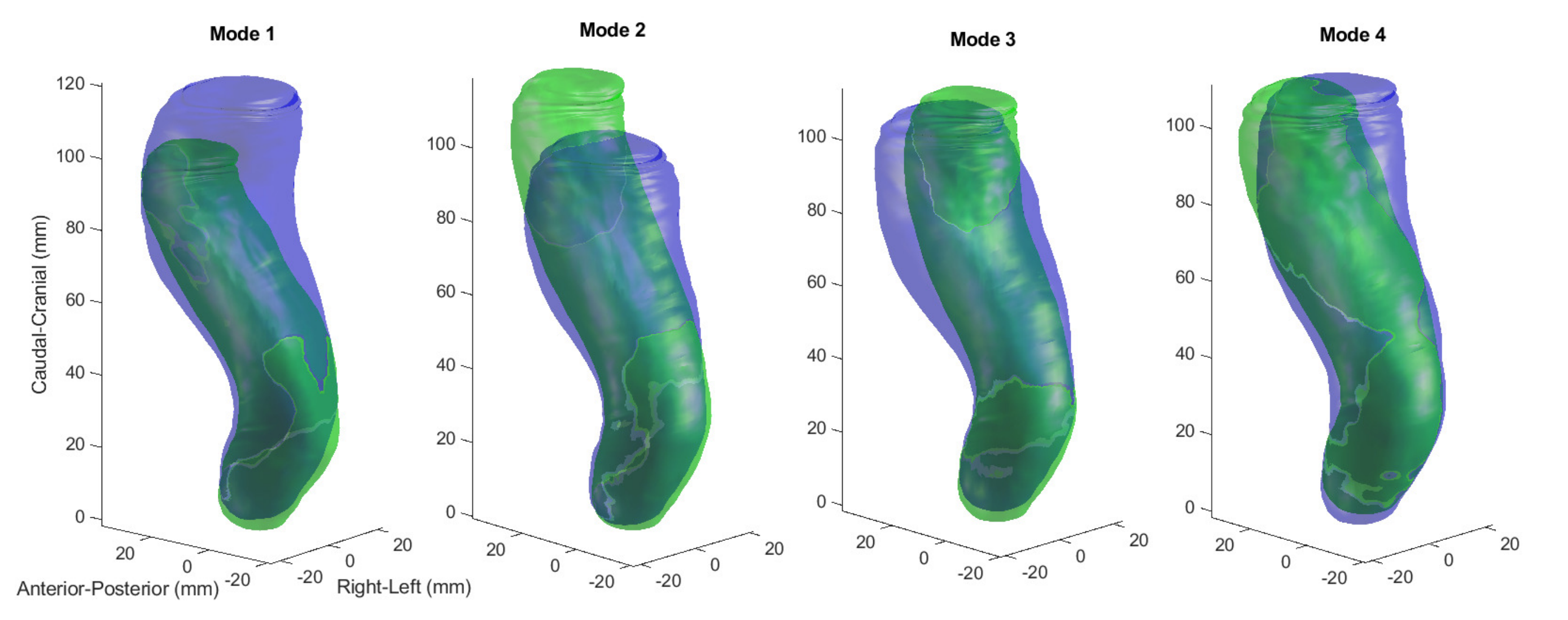}
    \caption{The four greatest population intra-patient deformation modes, shown with $\pm 2$ standard deviations from the population mean rectum. A video with animation of the deformation modes is available in the supplementary material [please insert link to video file here]}
    \label{fig:intramodes}
\end{figure*}
\begin{figure*}[h]
    \centering
    \includegraphics[width=0.95\textwidth]{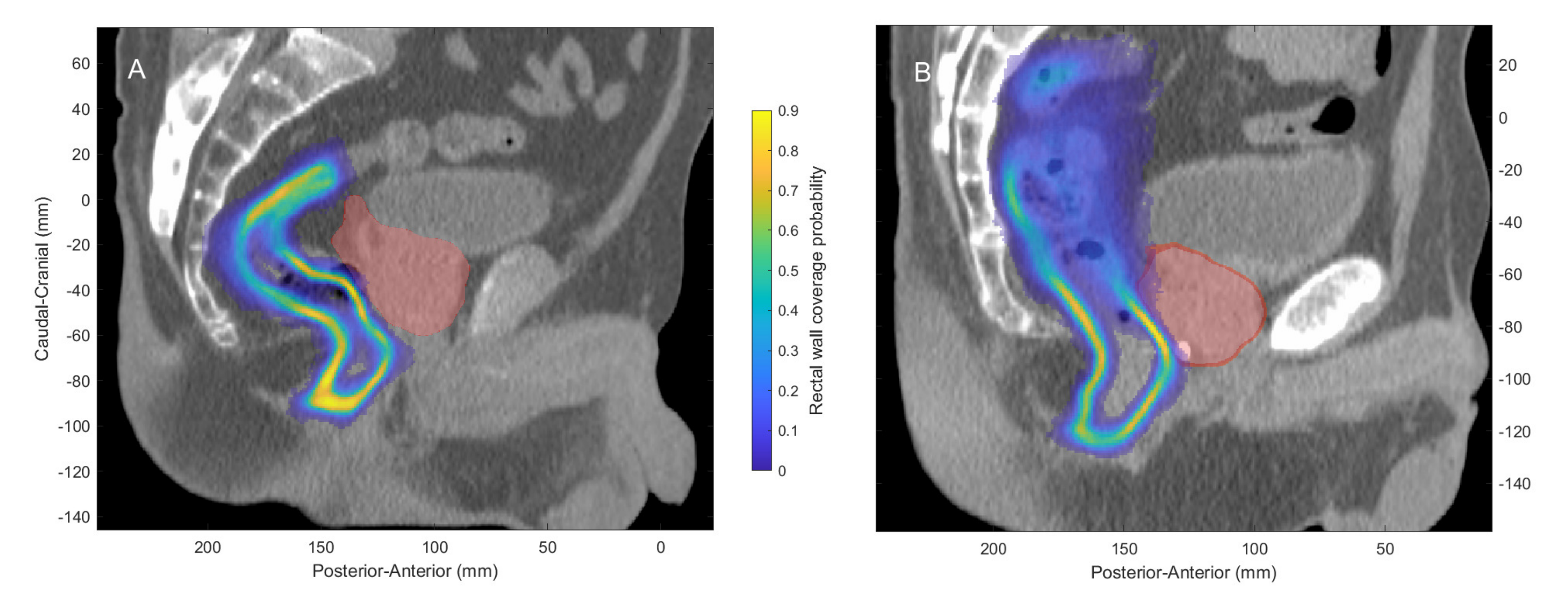}
    \caption{Coverage probability matrices for the rectal wall on a sagittal slice of the CT scan for two example patients --- a ``small mover'' (A) and a ``large mover'' (B). The red area is the high dose volume to the prostate, that receives more than 67 Gy EQD\textsubscript{2}}
    \label{fig:cpm}
\end{figure*}

\subsection{Impact of the uncertainty parameter \texorpdfstring{$\delta_R$}{delta R}}
For the variational Bayes model, the parameter $\delta_R$ naturally occured from the equations and the requirement that the covariance matrix must be non-singular. The PPCA method that we used to find $\delta_R$ can also be used for the other methods. We therefore tested the effect of $\delta_R$ on the the population model, the NIW model and the variational Bayes model, and compared the result to non-probabilistic PCA, i. e. $\delta_R = 0$. PPCA is not practical for the patient-specific model with as few as 3 input scans, since it requires that some principal components are not used. For the population model, $\delta_R$ was set constant, while for the NIW and variational method, it was updated according to the update equations for $\Psi$, which leads to
\begin{equation}
\delta_R(n) = \frac{n}{n+\nu}\delta_R(0),\label{deltaRn}
\end{equation}
where $n$ is the number of scans.

The motivation for this additional evaluation was to avoid a bias in favour of the variational Bayes model. 

\begin{figure*}[h]
    \centering
    \includegraphics[width=\textwidth]{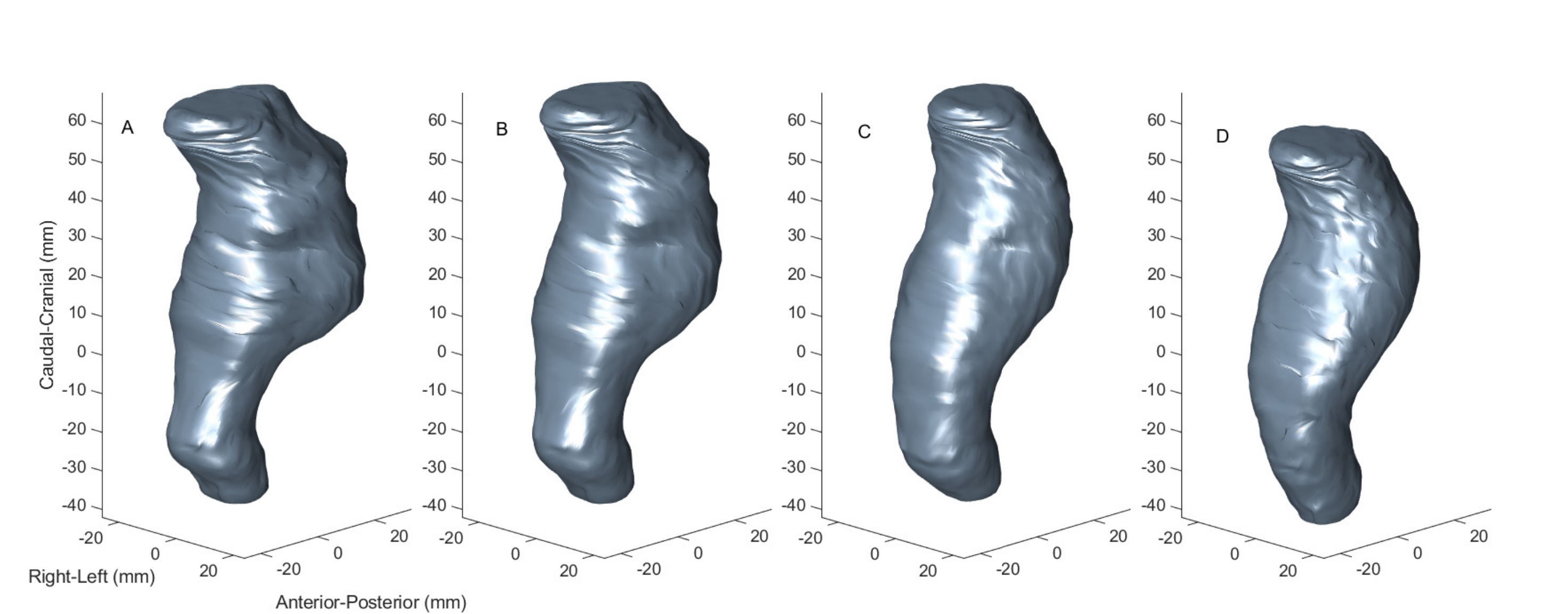}
    \caption{Estimated means and actual mean rectum shape for an example patient. A: Plan CT rectum shape, input to estimation algorithms. B: Mean rectum shape estimated by NIW model. C: Mean rectum shape estimated by variational Bayes model. D: Actual mean rectum shape over 9 scans.}
    \label{fig:meanrectums}
\end{figure*}
\begin{figure*}[h]
    \centering
    \includegraphics[width=\textwidth]{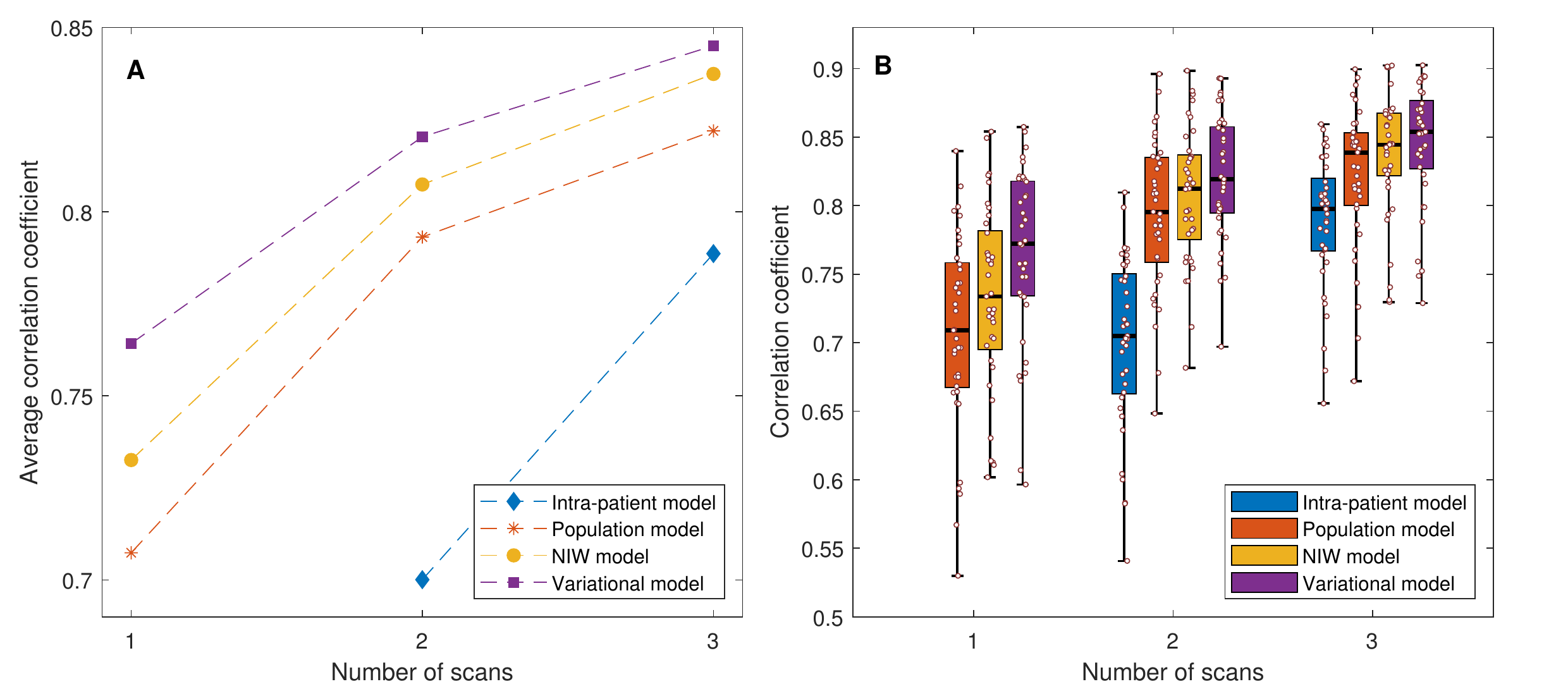}
    \caption{Correlation between the estimated CPMs and the references for the different methods using 1-3 input scans. A: Average correlation. B: Box plots showing median, 25th and 75th percentile and minimum and maximum values (whiskers). All individual values are also shown as circles over the box plot.}
    \label{fig:realdata}
\end{figure*}

\section{Results}
Visual comparison of the four first population intra-patient modes fits with anatomical expectations (Fig. \ref{fig:intramodes}). The first mode is mainly bending of the anorectal flexure; in the bent state, the rectum is less filled than in the straight state. The second mode shows stretching and compressing of the rectum in the caudal-cranial direction. The third mode shows mainly stretching of the top of the rectum in the left-right direction, while the fourth mode shows bending left-right of the top of the rectum. A general finding is that the most caudal third of the rectum, up to slightly above the anorectal flexure, moves very little. This is corroborated by figure \ref{fig:cpm}, which shows coverage probabilities of the rectum wall for two example patients, a ``small mover'' and a ``large mover''.

The Bayesian models take advantage of population data also when estimating the patient-specific mean rectum $\hat{\mu}$ . Figure \ref{fig:meanrectums} shows how the mean estimates may differ with the Bayesian models for an example patient, given a single input scan. For this patient, the mean shape from variational Bayes model had the greatest similarity with the true mean shape.

\begin{figure}[h]
    \centering
    \includegraphics[width=0.7\textwidth]{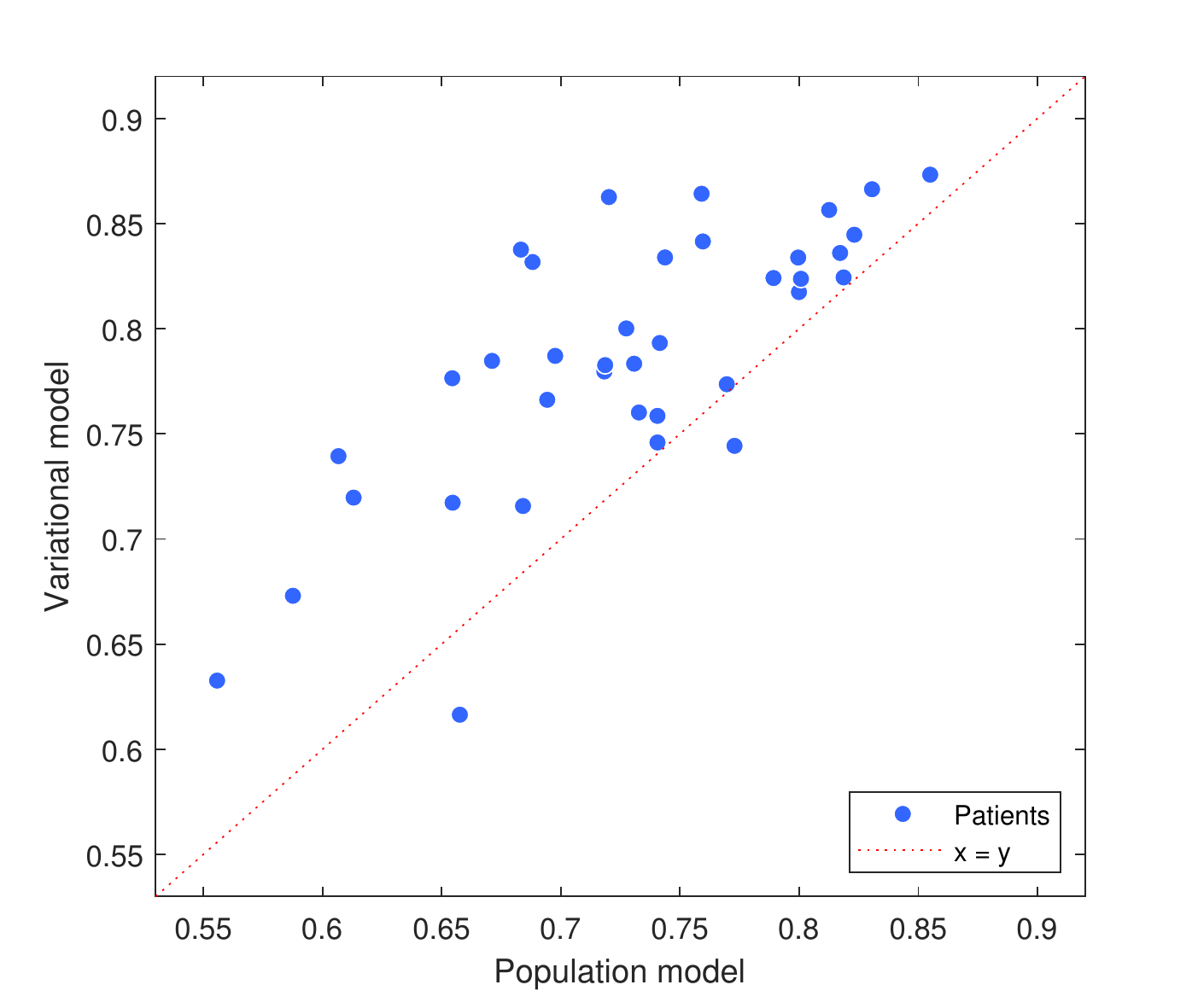}
    \caption{Correlation between the estimated CPMs and the references; comparison between the traditional population model and the proposed variational Bayes model. All points above the ``x=y'' line represent patients for which the variational Bayes method produced a better CPM estimate than the population model.}
    \label{fig:scatter}
\end{figure}

The average correlation between the estimated CPMs and the references is shown in Figure \ref{fig:realdata} A, while Figure \ref{fig:realdata} B shows the spread of the results among the individual patients. The two Bayesian methods outperform both the existing models, with the variational Bayes model showing superior results to the NIW-model. The results are summarized in table \ref{tab:results}, where the patient-specific model has been left out since it performs poorly with as few as three scans. The differences between the population, NIW and variational Bayes model were consistently significant (p \textless 0.05).
\begin{table*}[]
\caption{Difference in CPM correlation between the population, NIW and variational models using one, two and three scans. Here, $\Delta\mu$ is the difference in average value, and \%+ is the percentage of patients that saw improvement with the first method over the second.}
\label{tab:results}
\centering
\begin{tabular}{llll|lll|lll}
        & \multicolumn{3}{l|}{NIW vs pop. model}                                       & \multicolumn{3}{l}{Variational vs pop. model}          & \multicolumn{3}{l|}{Variational vs NIW}                \\ \cline{2-10} 
        & $\Delta \mu$              & p-value                     & \%+ & $\Delta \mu$               & p-value                     & \%+ & $\Delta \mu$              & p-value                     & \%+ \\ \cline{2-10} 
1 scan  & \multicolumn{1}{l|}{.026} & \multicolumn{1}{l|}{6.2e-5} & 78  & \multicolumn{1}{l|}{.058} & \multicolumn{1}{l|}{1.2e-8} & 95   & \multicolumn{1}{l|}{.032} & \multicolumn{1}{l|}{2.2e-6} & 81  \\
2 scans & \multicolumn{1}{l|}{.014} & \multicolumn{1}{l|}{1.8e-4} & 81  & \multicolumn{1}{l|}{.027}  & \multicolumn{1}{l|}{2.5e-6} & 86  & \multicolumn{1}{l|}{.013} & \multicolumn{1}{l|}{1.2e-3} & 70  \\
3 scans & \multicolumn{1}{l|}{.015} & \multicolumn{1}{l|}{2-2e-6} & 89  & \multicolumn{1}{l|}{.023}  & \multicolumn{1}{l|}{8.0e-7} & 89  & \multicolumn{1}{l|}{.008} & \multicolumn{1}{l|}{0.01}   & 62 
\end{tabular}
\end{table*}
In comparison to the best existing model (the population model), the variation Bayes model improved correlation with the reference CPM in 35 out of 37 patients when using a single input scan (Figure \ref{fig:scatter}).

\subsection{Convergence behaviour}
\begin{figure}[h]
    \centering
    \includegraphics[width=0.7\textwidth]{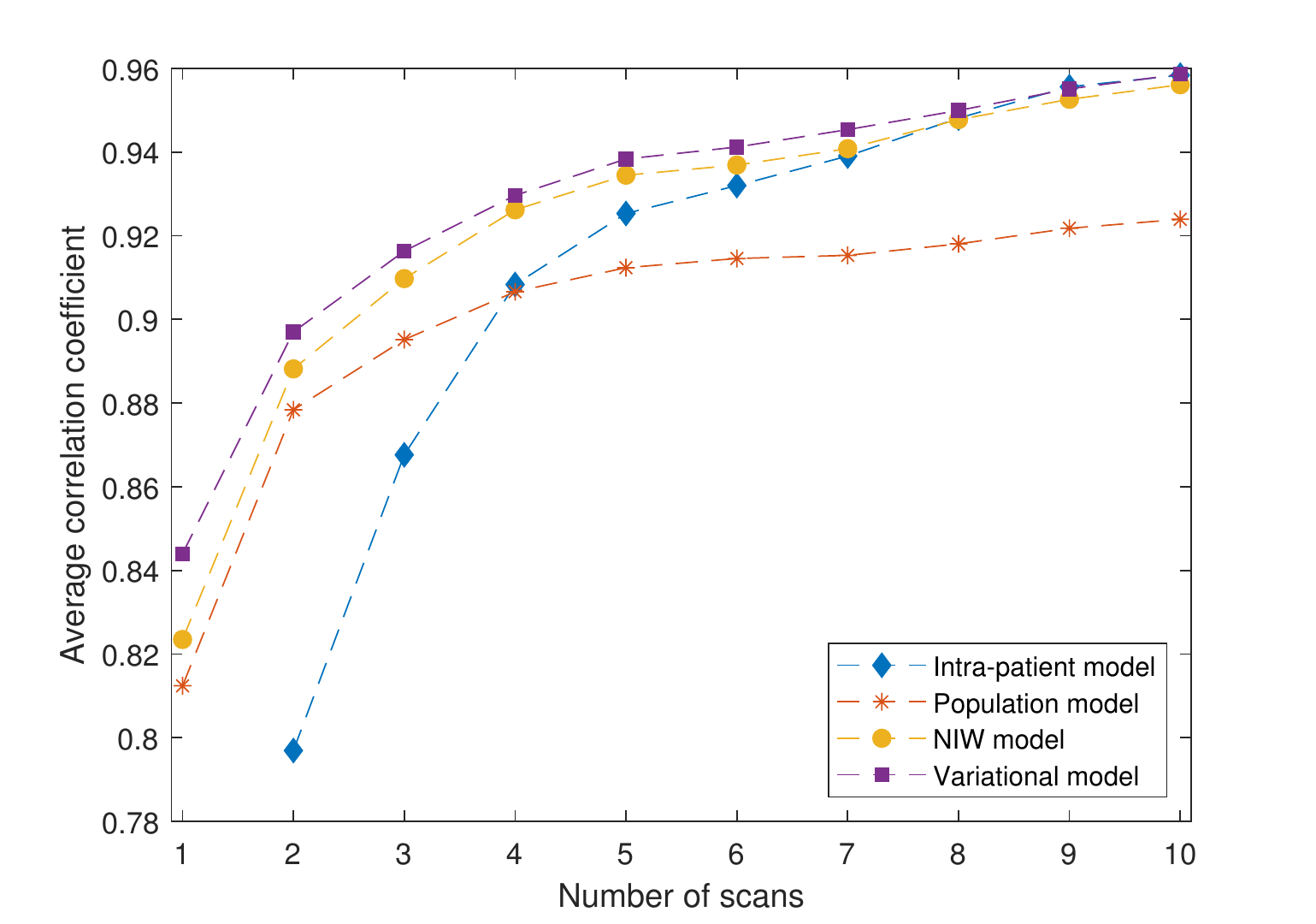}
    \caption{Average correlation between the estimated CPMs and the reference CPMs for the different methods using 1-10 scans, based on bootstrapped data.}
    \label{fig:simdata}
\end{figure}

The two Bayesian methods both outperform the patient-specific model with up to 6 scans, and outperforms the population model for any number of scans (Fig. \ref{fig:simdata}). As the number of input scans increases, the patient-specific model and the two Bayesian models appear to converge toward the true CPM, while the population model improves only moderately. This is to be expected, since, in the population model, the covariance matrix representing the random error is never updated. All improvement seen in the population model is therefore from reduction of error in the mean estimate, often referred to as systematic error. The performance of the patient-specific model is comparable to that of the population model when both are given 4 scans. For more than 4 scans, the patient-specific model outperforms the population model. The variational Bayes model consistently performs slightly better than the NIW-model.

\subsection{Impact of the uncertainty parameter \texorpdfstring{$\delta_R$}{delta R}}
\begin{figure}[h]
    \centering
    \includegraphics[width=0.7\textwidth]{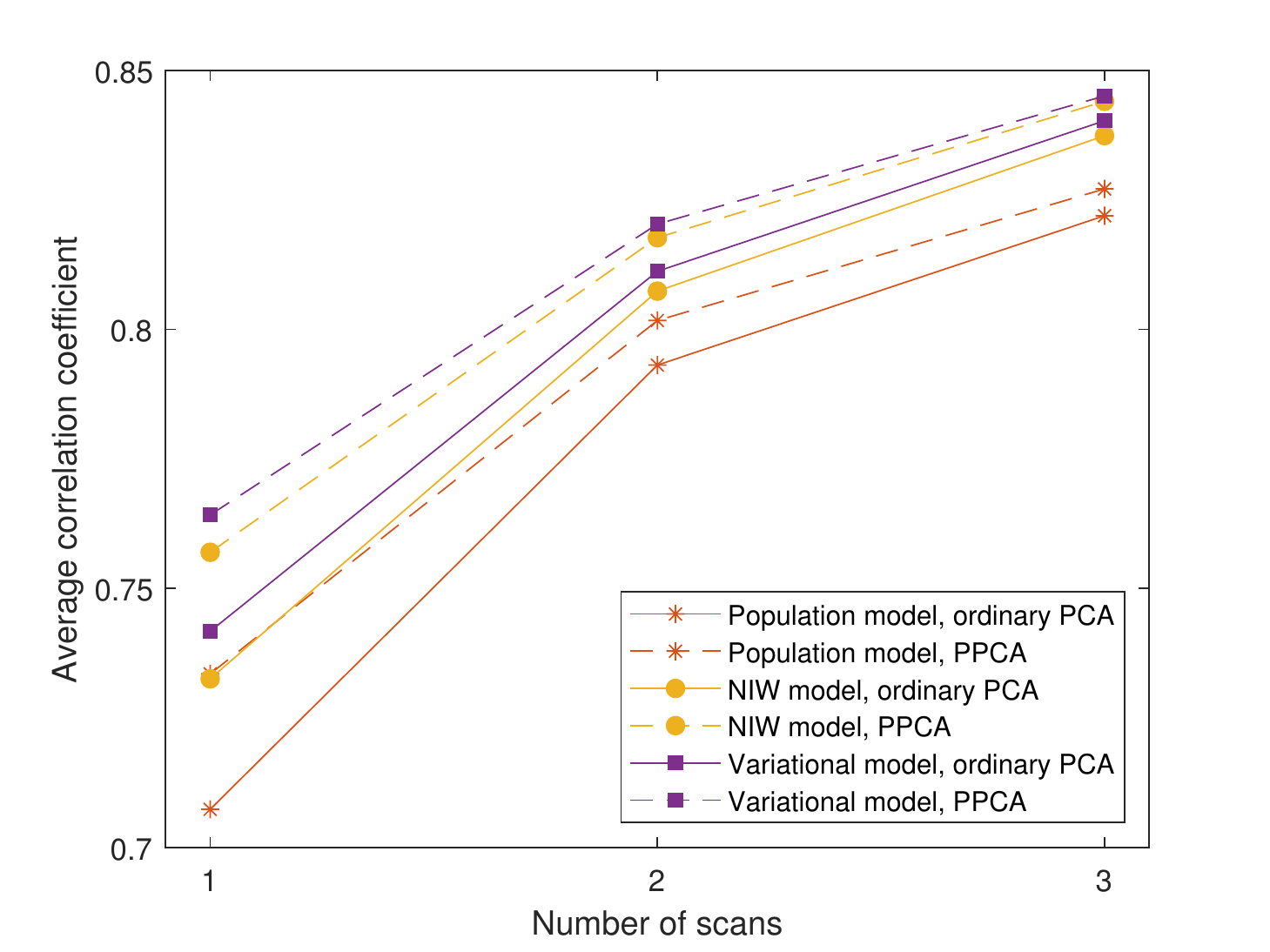}
    \caption{Comparison of ordinary and probabilistic PCA for different models. Each symbol represents the average correlation between the estimated CPMs and the references.}
    \label{fig:dRcomparison}
\end{figure}

   For all the models, PPCA through the addition of the $\delta_R$ parameter increases correlation as compared to ordinary PCA, as shown in Figure \ref{fig:dRcomparison}. The difference between the models with and without the uncertainty parameter is greatest when using a single scan. Although the differences between the models decreased, both Bayesian methods with ordinary PCA still perform the same as, or better than the population model with PPCA.

\section{Discussion}
Both the new models outperform the existing population model significantly. Conceptually, the NIW model is only slightly more complex than the population model, so there is little rationale for rather using the population model. Additionally, Figure \ref{fig:scatter} shows that the Bayesian models are robust, as evidenced by the fact that 35 out of 37 patients had improved result with the variational Bayes model over the population model  (29/37 for the NIW-model without PPCA). There is therefore very little risk involved in moving to a Bayesian model.

The choice between the two Bayesian methods is a tradeoff between model accuracy and complexity. The main concern with the variational Bayes model is the conceptual rather than the computational complexity --- it is more challenging to implement and requires more parameters than the NIW model. When using PPCA, the NIW models performance gets close to that of the variational Bayes model. 

As expected, the patient-specific model cannot compete with the other models when few scans are available. This model still has an advantage in that no training data from the population is required. Additionally, deformable registration is more readily available between contours of the same patient than between contours of different patients. There are therefore applications where the patient-specific model is the only available option. However, in these cases, care should be taken that sufficient scans are available, as shown in Figures \ref{fig:realdata} and \ref{fig:simdata}.

The convergence analysis in Figure \ref{fig:simdata} shows that we have achieved the goal of combining the advantages of both models; requiring few scans to achieve good accuracy while also improving accuracy with more scans. At around eight scans, the patient-specific model catches up with the Bayesian models. This is to be expected – at that point, the Bayesian models put very little weight on the population data since there is sufficient patient-specific data for an accurate model.

\subsection{Applications}
The calculation of CPMs play a key role in many applications of organ deformation models \citep{price_method_2007}.  The CPMs can be used for robust RT planning \citep{baum_robust_2006}, or to calculate margins based on the formula of \citet{stroom_inclusion_1999}, as in \citet{hysing_coverage_2011,thornqvist_treatment_2013,magallon-baro_modeling_2019}. In \citet{ramlov_clinical_2017,lindegaard_early_2017}, CPMs were used clinically to reduce toxicity in nodal boosting of cervical cancer RT.
Applications besides CPMs include robust evaluation through treatment course simulation \citep{sohn_dosimetric_2012, hysing_statistical_2018}, generation of plan libraries for RT personalized to motion \citep{rigaud_statistical_2019} and motion-robust optimization \citep{sobotta_robust_2010, unkelbach_robust_2018}. Recently, \cite{owens_development_2022} used a pure inter-patient model to reconstruct colorectal dose in childhood cancer survivors who had received RT with no CT simulation. Thus, applications also extends to improving evaluation of complications from RT.

The Bayesian approach offers additional advantages because it quantifies the model uncertainty. Consider for example the robust evaluation in \cite{sohn_dosimetric_2012, hysing_statistical_2018}: predicting dose-volume histograms (DVHs) with uncertainties (such as 5th and 95th percentiles). When using a non-Bayesian deformation model, the correctness of the predicted values rely on the correctness of the model's parameters. With a Bayesian model, the uncertainty of the parameters will translate to additional uncertainty regarding the dose-volume histogram, thus increasing the difference between the expected value and the 5/95 percentiles.

 Interfractional geometrical errors in RT are often divided into systematic and random errors. The random error is the motion around the mean shape and position at each fraction, while the systematic error is the difference between the actual mean and the estimated mean, usually the shape and position at the plan CT. In terms of the deformation models, the systematic error is the difference between the estimated and the true patient mean, $\hat{\mu}-\mu$. The presented Bayesian models reduces the systematic error as compared to the previous methods by utilizing population data when estimating $\hat{\mu}$ (see Figure \ref{fig:meanrectums}). In addition, the new models provide a personalized \emph{distribution} for the systematic error in terms of the posterior inter-patient distribution. The widely applied margin recipe by \cite{van_herk_probability_2000} uses the formula $2.5\Sigma+0.7\sigma$, where $\Sigma$ and $\sigma$ are the standard deviations of the systematic and random errors, respectively. Because the distribution of both the systematic and random errors are modeled under the Bayesian framework, it is in principle possible to use similar recipes for margins due to deformation.

\subsection{Choice of evaluation metric}
The cross-correlation metric puts proportionally higher weight on voxels that have a high coverage probability. Since a large portion of the organ tends to overlap in most or all shapes for one patient, all methods will tend to produce relatively high correlation values. Therefore, the differences between the methods may seem small. We still choose to use this metric because of its simplicity and ease of reproduction.

%\subsection{The scale of \texorpdfstring{$\delta_\Psi$}{delta-Psi} and \texorpdfstring{$\delta_\Lambda$}{delta-Lambda}}
\subsection{Parameter values}
The values of the scalar parameters in Table \ref{table:parameters} were all hand tuned. It is possible to optimize for these values (with the exception of $K$-inter and $K$-intra, which must be integers), using general optimization software. The cost function will depend on the application.
The parameter $\kappa$ for the NIW-model was set to 0.25. Using equation \eqref{muUpdateNIW}, we find that, given one input scan, this represents an \emph{shrinkage factor} of 0.2; i.e the estimated mean is “shrunk” by a factor 0.2 towards the population mean \citep{rortveit_reducing_2021}. The parameter $\nu$, the number of degrees of freedom of the Wishart distribution, was set to 6 for both the NIW and the variational model. Normally, $\nu$ represents the number of samples from which $\Psi$ was computed. However, this is under the assumption that these samples were all drawn from the same multivariate Gaussian distribution. In our case, the samples were drawn from $M$ different Gaussian distributions with covariance matrices $R_i$, none of which match a future patient's covariance matrix. Therefore, we are much less certain about $R$, and we need to choose a value for $\nu$ that is much smaller than the total number of observations in the training data.  

When tuning the values of $\delta_\Psi$ and $\delta_\Lambda$, we found that these needed to be set surprisingly large to achieve satisfactory results. Possibly, some assumptions or parts of the model do not actually fit the data well, and increasing the regularization values then compensates for the poor fit. This underscores the importance of evaluating the models with realistic data, and tailoring the parameters to the case at hand.

\subsection{Degenerate inverse Wishart distribution}
The inverse Wishart distribution is usually defined in terms of the (forward) Wishart distribution: If a random $n\times n$ matrix $G$ is Wishart distributed with $G\sim \mathcal{W}(\Psi, \nu)$, then its inverse $G^{-1}$ is inverse Wishart distributed with $G^{-1}\sim \mathcal{IW}(\Psi^{-1}, \nu)$. However, when $\nu < n$, the Wishart distribution is degenerate, as any matrix $G$ with a non-zero probability density has rank $\nu$ and is therefore singular. Then this definition of the IW distribution does not work. A singular inverse Wishart distribution is defined through the pseudo-inverse of $W$ (\cite{cook_mean_2011}, \cite{bodnar_singular_2016}). Unfortunately, this distribution is not well behaved, and does not have a finite expected value. Since we do not explicitly use the distribution, but rather a point estimate, this does not make a difference when using the models as described in this paper. However, care must be taken if using the full Bayesian model as described in section \ref{sec:extensions}, as individual realizations of $G$ can have very large eigenvalues.

\subsection{Extensions}\label{sec:extensions}
We have applied the models to the rectum alone, however, for use in e.g. robust optimization, it would be advantageous to model several structures simultaneously so that the correlation between structures are taken into account.

In the evaluation of the algorithms, we used point estimates for $\mu$ and $R$ as opposed to a full distribution. We have thus ignored the uncertainty in the model itself, and therefore sinned against the Bayesian philosophy. We chose to do this for the sake of computational complexity. However, it is possible to account for the additional uncertainty: When performing Monte-Carlo sampling, one would first sample $\mu$ and $R$ from the posterior distribution every time before sampling $s$ from $\mathcal{N}(\mu, R)$. The resulting distribution of $s$ is called the \emph{posterior predictive} distribution. Particularly the sampling of $R$ is computationally intensive. An alternative approach might therefore be to use a point estimate for $R$ while sampling $\mu$, as systematic errors are often of greater importance than random errors. 

The presented models have been applied to deformably registered organ surfaces. A more common form of deformable registration is the deformation of 3D-images with image intensities. Since both types of registration produce deformation vector fields, it is possible, with some adaptions, to apply these models to deformed images as well.

\section{Conclusions}
We have implemented and evaluated two Bayesian methods for modelling organ deformation occuring during RT treatment. The NIW and the variational Bayes models both outperformed previous organ deformation models when applied to the rectal wall of prostate cancer patients.

\anon{}{
%Tar dette inn igjen til slutt
%\section{Acknowledgements}
%The authors would like to thank Markus Alber for his contribution with ideas and discussions in the early phase of this project. We also thank Andras Zolnay at Erasmus MC Cancer center for valuable insight and discussion.
}

\anon{}{
\section{Funding}
This work was funded by Trond Mohn Foundation [grant number BFS2017TMT07].
}

%%Harvard
\bibliographystyle{model2-names}%\biboptions{authoryear}

\bibliography{references}

\begin{appendices}
\numberwithin{equation}{section}
\appendixpage
\section{Derivation of the conditional posteriors}\label{app:wishartnormal}
The pdf for the multivariate Gaussian distribution for a vector $x$ of dimension $p$ is
\begin{equation}\label{normaldist}
   \mathcal{N}(x;\mu,R)=\frac{1}{\sqrt{(2 \pi)^p |R|}}\exp (-\frac{1}{2}(x-\mu)^TR^{-1}(x-\mu)).
\end{equation}
The pdf of the inverse Wishart distribution of a $p\times p$ matrix $Q$ is
\begin{equation}\label{wishartdist}
\mathcal{IW}(Q;\Psi, \nu)=\frac{|\Psi|^{\nu/2}}{2^{\nu p/2}\Gamma_p(\frac{\nu}{2})}|Q|^{-(\nu+p+1)/2}\exp (-\frac{1}{2}\mrm{tr}(\Psi Q^{-1})).
\end{equation}
The joint pdf of $\mu$, $R$ and the samples $S=\{s_1,s_2,\ldots, s_n\}$, based on our prior and our likelihood is
\begin{equation}
  f(\mu, R,S)= \mathcal{N}(\mu;\mu_0, \Lambda)\mathcal{IW}(R;\Psi, \nu)\prod_{i=1}^n \mathcal{N}(s_i;\mu, R).
\end{equation}
Writing this out using \eqref{normaldist} and \eqref{wishartdist}, and leaving out any constant factors (factors that do not contain $\mu$, $R$ or $S$), we find
\begin{align}
    f(\mu, R,S)&\propto \frac{|R|^{-(\nu+p+1)/2}}{|R|^{n/2}}\exp \left(-\frac{1}{2}(\mu-\mu_0)^T\Lambda^{-1}(\mu-\mu_0)\right.\nonumber\\
    &\left.-\frac{1}{2}\mrm{tr}(\Psi R^{-1})-\frac{1}{2}\sum_{i=1}^n(s_i-\mu)^T R^{-1}(s_i-\mu)\right).\label{jointall}
\end{align}
Using the property of the trace $\mrm{tr}(ABC)=\mrm{tr}(CAB)$ and the fact that a scalar is its own trace, the sum within the exponential can be written as
\begin{equation}
\sum_{i=1}^n(s_i-\mu)^T R^{-1}(s_i-\mu)=\mrm{tr}\left([\sum_{i=1}^n(s_i-\mu)(s_i-\mu)^T]R^{-1}\right).
\end{equation}
Furthermore, since $\mrm{tr}(A)+\mrm{tr}(B)=\mrm{tr}(A+B)$, we can write
\begin{align}
    &\mrm{tr}(\Psi R^{-1})+\sum_{i=1}^n(s_i-\mu)^T R^{-1}(s_i-\mu)\nonumber\\
    &=\mrm{tr}([\Psi+\sum_{i=1}^n(s_i-\mu)(s_i-\mu)^T]R^{-1})\label{tracething}
\end{align}
To condition \eqref{jointall} on $\mu$ and $S$, we can leave out any factors not containing $R$ - that is, the first term in the exponenial. Using \eqref{tracething}, we find
\begin{align}
    f(R&|\mu, S)\nonumber\\
    &\propto |R|^{-(\nu+p+1+n)/2} \exp \left(-\frac{1}{2}\mrm{tr}([\Psi+\sum_{i=1}^n(s_i-\mu)(s_i-\mu)^T]R^{-1})\right)\nonumber\\
    &\propto \mathcal{IW}(R;\Psi^\prime, \nu^\prime),\label{app:fRcond1}
\end{align}
with
\begin{equation}
    \Psi^\prime =\Psi+\sum_{i=1}^n(s_i-\mu)(s_i-\mu)^T
\end{equation}
and
\begin{equation}
    \nu^\prime = \nu + n,
\end{equation}
which concludes the derivation of the conditional posterior for $R$.

Next, we condition \eqref{jointall} on $R$ and $S$ to find
\begin{align}
    &f(\mu|R, S)\nonumber\\
    &\propto \exp(-\frac{1}{2}(\mu-\mu_0)^T\Lambda^{-1}(\mu-\mu_0)-\frac{1}{2}\sum_{i=1}^n(s_i-\mu)^T R^{-1}(s_i-\mu))\label{app:fmufull}
\end{align}
Looking actively for a Gaussian distribution, we want to find that the terms inside the exponential are equal to 
\begin{equation}\label{desiredterms}
    -\frac{1}{2}(\mu-\mu_0^\prime)^T\Lambda^{\prime -1}(\mu-\mu_0^\prime) + c,
\end{equation}
for some $\Lambda^\prime$ and $\mu^\prime$, with any constant term $c$.
Grouping the terms that are quadratic in $\mu$, we find
\begin{equation}
    -\frac{1}{2}\mu^T\Lambda^{-1}\mu-\frac{1}{2}\sum_{i=1}^n\mu^T R^{-1} \mu = -\frac{1}{2}\mu^T(\Lambda^{-1}+nR^{-1})\mu,
\end{equation}
therefore, if this is a Gaussian distribution, we must have
\begin{equation}\label{Lambdaprimeapp}
    \Lambda^{\prime-1} = \Lambda^{-1}+nR^{-1}\rightarrow \Lambda^\prime=(\Lambda^{-1}+nR^{-1})^{-1}.
\end{equation}
Grouping the linear terms, we find
\begin{equation}
    -\mu^T\Lambda^{-1}\mu_0-\sum_{i=1}^n \mu^T R^{-1}s_i=-\mu^T(\Lambda^{-1}\mu_0+nR^{-1}\bar{s}).
\end{equation}
Setting this equal to the linear terms in \eqref{desiredterms}, we have
\begin{equation}
    \mu^T\Lambda^{\prime-1}\mu_0^{\prime}=\mu^T(\Lambda^{-1}\mu_0+nR^{-1}\bar{s}),
\end{equation}
which is true for any $\mu$ if and only if
\begin{equation}\label{mu0primeapp}
    \mu_0^\prime=\Lambda^\prime(\Lambda^{-1}\mu_0+nR^{-1}\bar{s}).
\end{equation}
The constant terms can be ignored, as they will be absorbed by the normalization. Finally, this gives us
\begin{align}
 f(\mu|R, S)\propto & \exp(-\frac{1}{2}(\mu-\mu_0^\prime)^T\Lambda^{\prime -1}(\mu-\mu_0^\prime))\\
 \propto &\quad \mathcal{N}(\mu;\mu_0^\prime, \Lambda^\prime),
\end{align}
with $\mu_0^\prime$ as in \eqref{mu0primeapp} and $\Lambda^\prime$ as in \eqref{Lambdaprimeapp}.
\qed

\section{Variational approximation}\label{app:variational}
To find the functions $q_\mu$ and $q_R$, we follow the procedure presented in \cite{gelman_bayesian_1995}. The minimizing functions are given by
\begin{equation}\label{app:qmu1}
    \log q_\mu(\mu)=E_{R}[\log f(\mu|R, S)]+\mrm{const}
\end{equation}
and
\begin{equation}\label{app:qR1}
    \log q_R(R)=E_{\mu}[\log f(R| \mu, S)] + \mrm{const},
\end{equation}
where $E_{R}$ and $E_{\mu}$ indicate an average over $R$ only or $\mu$ only, respectively. 

Inserting \eqref{app:fmufull} into \eqref{app:qmu1}, we get
\begin{align}
    \log q_\mu(\mu)&=E_R\bigg[-\frac{1}{2}(\mu-\mu_0)^T\Lambda^{-1}(\mu-\mu_0)\nonumber\\
    &-\frac{1}{2}\sum_{i=1}^n(s_i-\mu)^T R^{-1}(s_i-\mu)\bigg]+\mrm{const}\\
    =&-\frac{1}{2}(\mu-\mu_0)^T\Lambda^{-1}(\mu-\mu_0)\nonumber\\
    &-\frac{1}{2}\sum_{i=1}^n(s_i-\mu)^T E[R^{-1}](s_i-\mu)+\mrm{const}.
\end{align}
Following the lines of the derivation in appendix \ref{app:wishartnormal}, we find
\begin{equation}\label{app:qmu}
    q_\mu(\mu) = \mathcal{N}(\mu;\mu_0^*, \Lambda^*),
\end{equation}
with 
\begin{equation}\label{app:mustarex}
    \mu_0^*=(\Lambda^{-1}+nE[R^{-1}])^{-1}(\Lambda^{-1}\mu_0+nE[R^{-1}]\bar{s}).
\end{equation}
and
\begin{equation}\label{app:lambdastarex}
\Lambda^*=(\Lambda^{-1}+nE[R^{-1}])^{-1}.
\end{equation}
Similarly, we insert \eqref{app:fRcond1} into \eqref{app:qR1} to find
\begin{align}
    \log q_R(R) = &E_\mu\bigg[\mrm{log}(|R|^{-(\nu+p+1+n)/2})\nonumber\\ &-\frac{1}{2}\mrm{tr}([\Psi+\sum_{i=1}^n(s_i-\mu)(s_i-\mu)^T]R^{-1})\bigg] + \mrm{const}\\
    =& \mrm{log}(|R|^{-(\nu+p+1+n)/2})\nonumber\\
    &-\frac{1}{2}\mrm{tr}([\Psi+\sum_{i=1}^nE\left[(s_i-\mu)(s_i-\mu)^T\right]]R^{-1})+ \mrm{const}
\end{align}
The term within the expectation operator is
\begin{align}
    &\sum_{i=1}^n E[(s_i-\mu)(s_i-\mu)^T]\nonumber\\
    &=\sum_{i=1}^n \left(s_is_i^T+E[\mu \mu^T]-E[\mu]s_i^T-s_i E[\mu]^T\right)\\
    &=\sum_{i=1}^n (s_i-E[\mu])(s_i-E[\mu])^T+n(E[\mu \mu^T]-E[\mu]E[\mu]^T)\\
    &=\sum_{i=1}^n (s_i-E[\mu])(s_i-E[\mu])^T+n\cdot\mrm{cov}(\mu).\\
\end{align}
This leads to 
\begin{equation}\label{app:qR}
    q_R(R) = \mathcal{IW}(R;\Psi^*, \nu^*),
\end{equation}
with 
\begin{equation}
    \nu^* = \nu + n
\end{equation}
and
\begin{equation}\label{app:psistarex}
    \Psi^* = \Psi + \sum_{i=1}^n (s_i-E[\mu])(s_i-E[\mu])^T+n\cdot\mrm{cov}(\mu).
\end{equation}
Finally, we replace the moments in \eqref{app:mustarex}, \eqref{app:lambdastarex} and \eqref{app:psistarex} by the moments from the approximate distributions $q_\mu$ and $q_R$. Since $R$, according to $\eqref{app:qR}$, is inverse-Wishart distributed with scale matrix $\Psi^*$ and $\nu^*=\nu + n$ degrees of freedom, its inverse $R^{-1}$ is Wishart-distributed with scale matrix $\Psi^{*-1}$ and $\nu+n$ degrees of freedom. Its expectation is $E[R^{-1}] = \nu^*\Psi^{*-1}$. Therefore we find
\begin{equation}\label{app:mustarrep}
    \mu_0^*=(\Lambda^{-1}+n(\nu+n) \Psi^{*-1})^{-1}(\Lambda^{-1}\mu_0+n(\nu+n) \Psi^{*-1}\bar{s})
\end{equation}

and
\begin{equation}
\Lambda^*=(\Lambda^{-1}+n(\nu+n)\Psi^{*-1})^{-1}.   
\end{equation}
By \eqref{app:qmu}, the mean and covariace of $\mu$ is $\mu_0^*$ and $\Lambda^*$, therefore \eqref{app:psistarex} becomes
\begin{equation}
\Psi^* = \Psi + \sum_{i=1}^n (s_i-\mu_0^*)(s_i-\mu_0^*)^T+n\Lambda^*.    
\end{equation}

\section{Efficient computation of the update iteration}\label{app:computation}
The key to finding the estimated mean and covariance matrix for a patient is iteration over the update equations, repeated here for convenience:
\begin{align}
	\Lambda^*&=(\Lambda^{-1}+n(\nu+n)\Psi^{*-1})^{-1}\label{app:Lambdaupdate}\\
	\mu_0^*&=\Lambda^*(\Lambda^{-1}\mu_0+n(\nu+n)\Psi^{*-1}\bar{s})\label{app:muUpdate}\\
	\Psi^*& = \Psi + \sum_{j=1}^n (s_j-\mu_0^*)(s_j-\mu_0^*)^T+n\Lambda^* \label{app:PsiUpdate}\\
	\nu^*&=\nu+n
\end{align}
Only $\nu^*$ can be calculated directly. The other parameters rely on each other, and therefore require an iteration to converge to the correct values. 

Putting the iteration number $i$ in a superscript (replacing $\cdot^*$), we can write the iteration as
\begin{align}
	\Lambda^{(i)}&=(\Lambda^{-1}+n(\nu+n)\Psi^{(i-1)-1})^{-1}\label{LambdaupdateI}\\
\mu_0^{(i)}&=\Lambda^{(i)}(\Lambda^{-1}\mu_0+n(\nu+n)\Psi^{{(i-1)}-1}\bar{s})\label{muUpdateI}\\
\Psi^{(i)}& = \Psi + \sum_{j=1}^n (s_j-\mu_0^{(i)})(s_j-\mu_0^{(i)})^T+n\Lambda^{(i)} \label{PsiUpdateI}
\end{align}
We can see that we need to supply a starting guess for the first value $\Psi^{(0)}$.\footnote{Given that the iteration starts with the equation for $\Lambda^{(1)}$. If we had started with one of the other equations, a starting guess for at least one other parameter would need to be provided.} A natural starting guess is $\Psi^{(0)} = \Psi$.

In theory, the iteration represented by equations \eqref{LambdaupdateI}-\eqref{PsiUpdateI} can be implemented directly in any numerically oriented programming language. However, this would require storing and inverting very large $P\times P$ matrices, which is not attainable in practice. However, due to the structure of $\Psi$ and $\Lambda$ (when estimated as in sections \ref{psitraining} and \ref{lambdatraining}), memory and computation requirements can be drastically reduced.

Both matrices $\Lambda$ and $\Psi$ can be represented as an outer product of a data matrix with itself  plus a scalar multiple of the identity matrix:
\begin{align}
	\Lambda & = D_\Lambda D_\Lambda^T + \delta_\Lambda I\label{LambdaDef}\\
	\Psi & = D_\Psi D_\Psi^T + \delta_\Psi I.\label{PsiDef}
\end{align}
Here, $D_\Psi$ and $D_\Lambda$ are $P\times N_\Psi$ and $P\times N_\Lambda$ matrices, with $N_\Lambda,N_\Psi \ll P$. Multiplying a vector $a$ by such a matrix is much faster than the general $O(P^2)$ figure, since e. g. 
\begin{equation}
\Lambda a = (D_\Lambda D_\Lambda^T + \delta_\Lambda I)a = D_\Lambda(D_\Lambda^T a) + \delta_\Lambda a,
\end{equation}
which is easily computed in $O(N_\Lambda P)$ time. Furthermore, it is also fast to solve an equation such as $\Lambda x = b$. 

Throughout this derivation we shall make heavy use of the following special case of the \emph{Woodbury matrix identity}, which holds for any matrices $A$ and $B$ and scalar $\delta$ as long as the involved inversions are possible:
\begin{equation}
	(\delta I+ABA^T)^{-1}=\delta^{-1}I-\delta^{-1}A(\delta B^{-1}+A^TA)^{-1}A^T.\label{matinvlemma}
\end{equation}
This means that the inverses of $\Lambda$ and $\Psi$ can also be written in the form $DCD^T+\delta I$ for some $D$, $C$ and $\delta$.

\subsection{Computing \texorpdfstring{$\Lambda^{(i)}$}{Lambda(i)}}
We shall show later that $\Psi^{(i)}$ can be written for any $i$ as
\begin{equation}
	\Psi^{(i)}=D^{(i)}G^{(i)}D^{(i)T}+\delta_\Psi^{(i)}I,\label{PsiIdef}
\end{equation}
for some $\delta_\Psi^{(i)}$ and $G^{(i)}$, and where 
\begin{equation}
	D^{(i)}=\left[D_\Lambda \quad D_\Psi^{(i)}\right]
\end{equation}
for some $D_\Psi^{(i)}$ of dimension $P\times (N_\Psi+n)$. 
Inserting \eqref{LambdaDef} and \eqref{PsiIdef} into \eqref{LambdaupdateI}, we get
\begin{equation}
	\Lambda^{(i)}=[(D_\Lambda D_\Lambda^T + \delta_\Lambda I)^{-1}+n(\nu+n)(D^{(i-1)}G^{(i-1)}D^{(i-1)T}+\delta_\Psi^{(i-1)}I)^{-1}]^{-1}.\label{LambdaFirst}
\end{equation}
Using the matrix inversion lemma \eqref{matinvlemma} on both the inner inverses of \eqref{LambdaFirst}, we get
\begin{align}
	\Lambda^{(i)}=\left[\right.                                           &\delta_\Lambda^{-1} I-\delta_\Lambda^{-1}D_\Lambda(\delta_\Lambda I+D_\Lambda^T D_\Lambda)^{-1}D_\Lambda^T+n(\nu+n)\delta_\Psi^{(i-1)-1} I\nonumber\\
	-&n(\nu+n)\delta_\Psi^{(i-1)-1}D^{(i-1)}(\delta_\Psi^{(i-1)} G^{(i-1)-1}+ D^{(i-1)T} D^{(i-1)})^{-1}D^{(i-1)T}]^{-1}\label{messyinverse2}
\end{align}
In order to group the terms, note that 
\begin{equation}
-\delta_\Lambda^{-1}D_\Lambda(\delta_\Lambda I+D_\Lambda^T D_\Lambda)^{-1}D_\Lambda^T=D^{(i-1)}QD^{(i-1)T},
\end{equation}
where Q is a block-diagonal matrix
\begin{equation}
	Q=\left[\begin{array}{cc}
		-\delta_\Lambda^{-1}(\delta_\Lambda I + D_\Lambda^T D_\Lambda)^{-1} & \\
		& 0_{N_\Psi+n\times N_\Psi+n}
	\end{array}\right].
\end{equation}
We also define
\begin{equation}\label{Lidef}
	L^{(i)} = -\delta_\Psi^{(i)-1}(\delta_\Psi^{(i)} G^{(i)-1}+ D^{(i)T} D^{(i)})^{-1}
\end{equation}
and
\begin{equation}
	F^{(i)} = Q+n(\nu+n)L^{(i-1)}
\end{equation}
Now, we can write
\begin{equation}
	\Lambda^{(i)}=(\delta_\Lambda^{-1} +n(\nu+n)\delta_\Psi^{(i-1)-1})I+D^{(i-1)}F^{(i)}D^{(i-1)T})^{-1},\label{Lambdainverse}
\end{equation}
Applying the matrix inversion lemma again, we find
\begin{equation}\label{finalinverse}
	\Lambda^{(i)}=\delta^{(i)} I-\delta^{(i)} D^{(i-1)}H^{(i)}D^{(i-1)T},
\end{equation}
where
\begin{equation}
	H^{(i)}=[D^{(i-1)T}D^{(i-1)}+\delta^{(i)-1} F^{(i)-1}]^{-1}
\end{equation}
and
\begin{equation}
	\delta^{(i)} = (\delta_\Lambda^{-1}+n(\nu+n)\delta_\Psi^{(i-1)-1})^{-1}.
\end{equation}
Equation \eqref{finalinverse} gives us an expression for $\Lambda^{(i)}$ using only lower dimensional matrices and scalars. In practice, we never construct $\Lambda^{(i)}$ --- it is represented implicitly by $D^{(i)}$, $H^{(i)}$ and $\delta^{(i)}$ through \eqref{finalinverse}.

\subsection{Computing \texorpdfstring{$\mu_0^{(i)}$}{mu(i)}}
Through the derivation of $\Lambda^{(i)}$, we have already come a long way towards computing $\mu_0^{(i)}$. We can write \eqref{muUpdateI} as
\begin{equation}\label{mu0r}
	\mu_0^{(i)}=\Lambda^{(i)}r^{(i)},
\end{equation}
with
\begin{equation}\label{rdef}
	r^{(i)}=\Lambda^{-1} \mu_0 + n(\nu+n)\Psi^{(i-1)-1}\bar{s}.
\end{equation}
The first term of \eqref{rdef} is constant, and can be computed once. Using the matrix inversion lemma on \eqref{LambdaDef}, we find
\begin{equation}
	\Lambda^{-1}\mu_0=\delta_\Lambda^{-1}\mu_0-\delta_\Lambda^{-1}D_\Lambda(\delta_\Lambda I + D_\Lambda^T D_\Lambda)^{-1}(D_\Lambda\mu_0)
\end{equation}
The last term needs to be computed for each iteration. We find it by using the matrix inversion lemma on \eqref{PsiIdef}:
\begin{align}
	\Psi^{(i)-1}\bar{s}&=(\delta_\Psi^{(i)-1}I-\delta_\Psi^{(i)-1}D^{(i)}(\delta_\Psi^{(i)}G^{(i)-1}+D^{(i)T}D^{(i)})^{-1}D^{(i)T})\bar{s}\\
	&=\delta_\Psi^{(i)-1}\bar{s}+D^{(i)}L^{(i)}(D^{(i)T}\bar{s})
\end{align}
Finally, inserting \eqref{finalinverse} into \eqref{mu0r}, we find
\begin{equation}
	\mu_0^{(i)}=\delta^{(i)}r^{(i)} -\delta^{(i)} D^{(i-1)}H^{(i)}(D^{(i-1)T}r^{(i)}).
\end{equation}
\subsection{Computing \texorpdfstring{$\Psi^{(i)}$}{Psi(i)}}
The update equation for $\Psi$ is
\begin{align}
	\Psi^{(i)} =& \Psi + \sum_{j=1}^n (s_j-\mu_0^{(i)})(s_j-\mu_0^{(i)})^T+n\Lambda^{(i)}\nonumber\\
	=& D_\Psi D_\Psi^T+\delta_\Psi I + \sum_{j=1}^n (s_j-\mu_0^{(i)})(s_j-\mu_0^{(i)})^T+n\Lambda^{(i)}.
\end{align}
We can augment the data matrix $D_\Psi$ by inserting new columns which are the mean-subtracted data vectors;
\begin{equation}\label{Dupdate}
	D^{(i)}_\Psi=[D_\Psi \quad s_1-\mu_0^{(i)} \quad s_2-\mu_0^{(i)}\quad\ldots\quad s_n-\mu_0^{(i)}],
\end{equation}
and we find
\begin{equation}
	\Psi^{(i)}=D_\Psi^{(i)} D_\Psi^{(i)T}+\delta_\Psi I+n\Lambda^{(i)}.
\end{equation}
Inserting \eqref{finalinverse}, we get
\begin{equation}
	\Psi^{(i)}=D_\Psi^{(i)} D_\Psi^{(i)T}+\delta_\Psi I+n(\delta^{(i)} I-\delta^{(i)} D^{(i-1)}H^{(i)}D^{(i-1)T})
\end{equation}
We want to group the terms of this equation, but run into a slight problem: One term contains $D_\Psi^{(i)}$, while another term contains $D^{(i-1)}$ (which contains $D_\Psi^{(i-1)})$. In practice, this can easily be resolved by replacing $D^{(i-1)}$ by $D^{(i)}$; this is in line with the algorithm philosophy of always using the most recent guess of each parameter, and also guarantees that the equations \eqref{app:Lambdaupdate}-\eqref{app:PsiUpdate} hold at convergence (at convergence, we have $D^{(i)}=D^{(i-1)}$).
Now, to group the terms, first note that 
\begin{equation}
	D_\Psi^{(i)} D_\Psi^{(i)T}=D^{(i)}KD^{(i)T},
\end{equation}
where
\begin{equation}
	K=\left[\begin{array}{cc}
		0_{N_\Lambda\times N_\Lambda}&  \\
		&  I_{N_\Psi+n} \end{array}	\right].
\end{equation}
Thus, we can write
\begin{equation}
	\Psi^{(i)}=D^{(i)} (K-n\delta^{(i)}H^{(i)})D^{(i)T}+(\delta_\Psi+n\delta^{(i)}) I.
\end{equation}
We now see that we must have
\begin{equation}
	G^{(i)}=K-n\delta^{(i)}H^{(i)}
\end{equation}
and 
\begin{equation}
	\delta_\Psi^{(i)}=\delta_\Psi+n\delta^{(i)}
\end{equation}
in order for $\Psi^{(i)}$ to be written as
\begin{equation}
	\Psi^{(i)}=D^{(i)}G^{(i)}D^{(i)T}+\delta_\Psi^{(i)}I.
\end{equation}
\subsection{Initial values}
Initially, we want to get $\Psi^{(0)}=\Psi$, i. e. $D^{(0)}G^{(0)}D^{(0)T}+\delta_\Psi^{(0)} I=D_\Psi D_\Psi^T+\delta_\Psi I$  which achieve by setting
\begin{align}
	\delta_\Psi^{(0)}&=\delta_\Psi\\
	D_\Psi^{(0)}&=[D_\Psi\quad 0_{P\times n}]\\
	G^{(0)}&=K.
\end{align}
However, $G^{(0)}$ is not invertible, which makes it impossible to compute $L^{(0)}$ as in \eqref{Lidef}. Instead, $L^{(0)}$ must be initialized to
\begin{equation}
	L^{(0)}=\left[\begin{array}{cc}0_{N_\Lambda \times N_\Lambda} &\\ & -\delta_\Psi^{-1}(\delta_\Psi I+ D_\Psi^{(i)T} D_\Psi^{(i)})^{-1}\end{array}\right].
\end{equation}

\subsection{Algorithm summary}
\begin{algorithmic}
	\State \textbf{Input:} $\mu_0$, $D_\Lambda$, $D_\Psi$, $\delta_\Psi$, $\delta_\Lambda$, $s_1 \ldots s_n$, $\nu$
	\State \textbf{Output:} $\mu_0^*$, $D^*$, $G^*$, $\delta_\Psi^*$, $\delta^*$, $H^*$
	\State
	\State $K=\left[\begin{array}{cc}
		0_{N_\Lambda\times N_\Lambda}&  \\
		&  I_{N_\Psi+n} \end{array}	\right].$
	\State $Q=\left[\begin{array}{cc}
		-\delta_\Lambda^{-1}(\delta_\Lambda I + D_\Lambda^T D_\Lambda)^{-1} & \\
		& 0_{N_\Psi+n\times N_\Psi+n}
	\end{array}\right]$
	\State $q \gets \delta_\Lambda^{-1}\mu_0-\delta_\Lambda^{-1}D_\Lambda(\delta_\Lambda I + D_\Lambda^T D_\Lambda)^{-1}(D_\Lambda^T\mu_0)$\quad /* $q$ is $\Lambda^{-1}\mu_0$ */
	\State $D_\Psi^{(0)}=[D_\Psi \quad 0_{P\times n}]$
	\State $D^{(0)} \gets [D_\Lambda\quad D_\Psi^{(0)}]$
	\State	$\delta_\Psi^{(0)}\gets\delta_\Psi$
	\State $\mu_0^{(0)} \gets \mu_0$
	\State $L^{(0)}\gets\left[\begin{array}{cc}0_{N_\Lambda \times N_\Lambda} &\\ & -\delta_\Psi^{-1}(\delta_\Psi I+ D_\Psi^{(0)T} D_\Psi^{(0)})^{-1}\end{array}\right]$
	\State $i\gets 0$
	
	\Repeat
	\State $i\gets i+1$
	\State $\delta^{(i)} \gets (\delta_\Lambda^{-1}+n(\nu+n)\delta_\Psi^{(i-1)-1})^{-1}$
	\State	$F^{(i)} \gets  Q+n(\nu+n)L^{(i-1)}$
	\State $H^{(i)}\gets (D^{(i-1)T}D^{(i-1)}+\delta^{(i)-1} F^{(i)-1})^{-1}$
	\State $r^{(i)}\gets q+n(\nu+n)(\delta_\Psi^{(i-1)-1} I+D^{(i-1)}L^{(i-1)}D^{(i-1)T})\bar{s}$
	\State $\mu_0^{(i)}\gets \delta^{(i)}r^{(i)} -\delta^{(i)} D^{(i-1)}H^{(i)}(D^{(i-1)T}r^{(i)})$
	\State $D^{(i)}_\Psi \gets [D_\Psi\quad s_1-\mu^{(i)}_0\quad s_2-\mu^{(i)}_0\quad\ldots\quad s_n-\mu^{(i)}_0]$
	\State $D^{(i)} \gets [D_\Lambda\quad D_\Psi^{(i)}]$
	\State	$\delta_\Psi^{(i)}\gets\delta_\Psi+n\delta^{(i)}$
	\State $G^{(i)}\gets K-n\delta^{(i)}H^{(i)}$
	\State $L^{(i)} \gets -\delta_\Psi^{(i)-1}(\delta_\Psi^{(i)} G^{(i)-1}+ D^{(i)T} D^{(i)})^{-1}$
	\Until{$\|\mu_0^{(i)}-\mu_0^{(i-1)}\|<\epsilon$}
	\State $\mu_0^*\gets \mu^{(i)}$, $D^*\gets D^{(i)}$, $G^*\gets G^{(i)}$, $\delta_\Psi^*\gets \delta_\Psi^{(i)}$, $\delta^*\gets \delta^{(i)}$, $H^*\gets H^{(i)}$
	\State
	\State /* Implicit, not computed: $\Lambda^{*}=\delta^{*} I-\delta^{*} D^{*}H^{*}D^{*T}$ */
	\State /* Implicit, not computed: $\Psi^{*}=D^{*}G^{*}D^{*T}+\delta_\Psi^{*}I$ */		
\end{algorithmic}

\section{Bias of the inter-patient covariance matrix estimate}
\label{app:bias}
We estimate the inter-patient covariance matrix as 
\begin{equation}
    \hat{\Lambda}=\frac{1}{M-1}\sum_{i=1}^M(\bar{s}_i-\hat{\mu}_0)(\bar{s}_i-\hat{\mu}_0)^T.\label{LambdaHat1}
\end{equation}
This is the sample covariance matrix of $\bar{s}_i$, as opposed to $\mu$ which we are interested in. But $\bar{s}_i$ are not identically distributed if $J_i$ varies. We can show that
\begin{equation}
E[\hat{\Lambda}] = \frac{1}{M}\sum_{i=1}^M\mrm{cov}(\bar{s}_i).\label{ELambdaHat1}
\end{equation}
To avoid clutter, the proof of this result is given at the end of the appendix.

The covariance matrix of a sample mean based on $n$  i.i.d. samples is always given by $1/n$ times the covariance matrix of one sample. In other words,
\begin{equation}
\mrm{cov}(\bar{s}_i|\mu,R) = \frac{1}{J_i}R.
\end{equation}
Now we can use the \emph{law of total covariance}, which states, for two scalar random variables $a$ and $b$, 
\begin{equation}
\mrm{cov}(a, b)=E[\mrm{cov}(a, b | c)]+\mrm{cov}(E[a|c],E[b|c]).
\end{equation}In our case, we get
\begin{align}
\mrm{cov}(\bar{s}_i)=&E[\mrm{cov}(\bar{s}_i|\mu, R)]+\mrm{cov}(E[\bar{s}_i|\mu, R])\\
=&E[\frac{1}{J_i}R]+\mrm{cov}(\mu)\\
=&\frac{1}{J_i}E[R]+\Lambda\label{covSi}
\end{align}
since, by definition, $\mrm{cov}(\mu)=\Lambda$. Inserting \eqref{covSi} into \eqref{ELambdaHat1} yields
\begin{equation}
E[\hat{\Lambda}]= \frac{1}{M}\sum_{i=1}^M\left(\Lambda + \frac{1}{J_i}E[R]\right) = \Lambda + cE[R],
\end{equation}
where 
\begin{equation}
    c = \frac{1}{M}\sum_{i=1}^M\frac{1}{J_i}.
\end{equation}
\qed

\textbf{Proof of \eqref{ELambdaHat1}:}

We start by manipulating \eqref{LambdaHat1}:
\begin{align}
    \hat{\Lambda}=&\frac{1}{M-1}\sum_{i=1}^M(\bar{s}_i-\hat{\mu}_0)(\bar{s}_i-\hat{\mu}_0)^T\\
    =&\frac{1}{M-1}\left(\sum_{i=1}^M\bar{s}_i\bar{s}_i^T+\sum_{i=1}^M\hat{\mu}_0\hat{\mu}_0^T-\sum_{i=1}^M\bar{s}_i\hat{\mu}_0^T-\sum_{i=1}^M\hat{\mu}_0\bar{s}_i^T\right)\\
    =&\frac{1}{M-1}\left(\sum_{i=1}^M\bar{s}_i\bar{s}_i^T+\sum_{i=1}^M\hat{\mu}_0\hat{\mu}_0^T-(\sum_{i=1}^M\bar{s}_i)\hat{\mu}_0^T-\hat{\mu}_0(\sum_{i=1}^M\bar{s}_i^T)\right)\\    
    =&\frac{1}{M-1}\left(\sum_{i=1}^M\bar{s}_i\bar{s}_i^T+M\hat{\mu}_0\hat{\mu}_0^T-M\hat{\mu}_0\hat{\mu}_0^T-M\hat{\mu}_0\hat{\mu}_0^T\right)\\
     =&\frac{1}{M-1}\left(\sum_{i=1}^M\bar{s}_i\bar{s}_i^T-M\hat{\mu}_0\hat{\mu}_0^T\right),   
\end{align}
where we used $\hat{\mu}_0=\frac{1}{M}\sum_{i=1}^M \bar{s}_i$. Taking the expectation, and using the general formula $E[xx^T]=\cov{x}+E[x]E[x]^T$, we find
\begin{align}
    E[\hat{\Lambda}]=&\frac{1}{M-1}\left(\sum_{i=1}^ME[\bar{s}_i\bar{s}_i^T]-ME[\hat{\mu}_0\hat{\mu}_0^T]\right)\\
    =&\frac{1}{M-1}\sum_{i=1}^M(\cov{\bar{s}_i}+E[\bar{s}_i]E[\bar{s}_i]^T)\nonumber\\
    &-\frac{M}{M-1}(\cov{\hat{\mu}_0}+E[\hat{\mu}_0]E[\hat{\mu}_0]^T)\\
    =&\frac{1}{M-1}\left(\sum_{i=1}^M\cov{\bar{s}_i}-M\cov{\hat{\mu}_0}\right),\label{ELambdahat3}
\end{align}
since $E[\bar{s}_i]=E[E[\bar{s}_i|\mu]]=E[\mu]=\mu_0=E[\hat{\mu}_0]$. Looking at $\cov{\hat{\mu}_0}$, we find
\begin{align}
    \cov{\hat{\mu}_0} =& \cov{\frac{1}{M}\sum_{i=1}^M\bar{s}_i}\\
    =&\frac{1}{M^2}\sum_{i=1}^M\cov{\bar{s}_i},\label{covmuhat}
\end{align}
since $\bar{s}_i$ are independent (though not identically distributed). Inserting \eqref{covmuhat} into \eqref{ELambdahat3} yields
\begin{align}
    E[\hat{\Lambda}]=&\frac{1}{M-1}\left(\sum_{i=1}^M\cov{\bar{s}_i}-\frac{M}{M^2}\sum_{i=1}^M\cov{\bar{s}_i}\right)\\
    =&\frac{1}{M-1}(1-\frac{1}{M})\sum_{i=1}^M\cov{\bar{s}_i}\\
    =&\frac{1}{M}\sum_{i=1}^M\cov{\bar{s}_i}.
\end{align}

\qed
\section{PCA for the bias-corrected inter-patient covariance matrix}
\label{app:PCAbiascorrected}
The bias-corrected inter-patient covariance matrix estimate is given by 
\begin{equation}
\tilde{\Lambda} = \hat{\Lambda}-c \hat{R}_\mrm{pop},
\end{equation}
where $c=\frac{1}{M}\sum_{i=1}^M\frac{1}{J_i}$. This matrix is not positive semidefinite, and cannot be expressed with a real-valued data matrix $D$ as $\tilde{\Lambda}=DD^T$. It can, however, be expressed as
\begin{equation}
\tilde{\Lambda} =AB^T,
\end{equation}
where  $A=[D_\Lambda\quad \sqrt{c}D_\mrm{pop}]$ and $B=[D_\Lambda\quad -\sqrt{c}D_\mrm{pop}]^T$. 

As usual $\tilde{\Lambda}$ is too big to practically perform eigenvalue decompostion on. However, there is a relation between the eigenvalue decomposition of $AB^T$ and that of $B^TA$ . The latter is a small matrix, and its eigenvalue decomposition can easily computed using any numerical software package. Given the $k$th eigenvalue $\lambda_k$ and the $k$th eigenvector $v_k$ of $B^TA$, the $k$th eigenvalue of $\tilde{\Lambda}$ is $\lambda_k$, and the $k$th eigenvector is
\begin{equation}\label{wkprime}
w_k = Av_k.
\end{equation}
A proof of this result is given at the end of the appendix. The scale of $w_k$ is arbitrary, so we want to normalize it as
\begin{equation}
w_k^\prime = \frac{w_k}{\|w_k\|}.
\end{equation}
As usual in PCA, we discard the eigenpairs corresponding to the smallest eigenvalues. In this case, since the matrix is not positive semidefinite, several of the eigenvalues will be negative. We need to discard all eigenpairs corresponding to negative eigenvalues, since we cannot have negative variance for any of the modes (which would lead to a complex data matrix).
The PCA-reduced covariance matrix can now be represented by a data matrix $\tilde{D}_\mrm{PCA}$ as
\begin{equation}
\tilde{\Lambda}_\mrm{PCA}=\tilde{D}_\mrm{PCA}\tilde{D}_\mrm{PCA}^T,
\end{equation}
with
\begin{equation}
\tilde{D}_\mrm{PCA}=\left[\sqrt{\lambda_1} w_1^\prime \quad \sqrt{\lambda_2} w_2^\prime \quad \ldots\quad \sqrt{\lambda_K} w_K^\prime\right].
\end{equation}
\textbf{Proof of \eqref{wkprime}:} \\
 Let $w_k$ be an eigenvector of $AB^T$, and $\lambda_k$ be the corresponding eigenvalue, i. e.
\begin{equation}
AB^Tw_k = \lambda_k w_k.
\end{equation}
We can transform $AB^T$ into $B^TA$ by what we may call a pseudo-similarity transformation:
\begin{equation}
A^+(AB^T)A=(A^TA)^{-1}A^TAB^TA=B^TA,
\end{equation}
where $A^+$ denotes the pseudo-inverse of $A$. Also note that $AA^+$ is a projection matrix onto the subspace spanned by $A$. Since $w_k$, as an eigenvector of $AB^T$, is already in this subspace, we have
\begin{equation}
AA^+w_k = w_k.\label{wprojection}
\end{equation}Using the three previous equations, we can now write
\begin{equation}
B^TA(A^+w_k)=A^+AB^TAA^+w_k=A^+AB^Tw_k=\lambda_kA^+w_k.
\end{equation}This shows that $\lambda_k$ is an eigenvalue of $B^TA$, with corresponding eigenvector $v_k=A^+w_k$. However, we want to find $w_k$ given $v_k$. Using \eqref{wprojection} again, we find
\begin{align}
v_k&=A^+w_k\\
\rightarrow Av_k&=AA^+w_k=w_k.
\end{align}
\qed

\end{appendices}

\end{document}